\documentclass[prd,twocolumn,nofootinbib,showpacs,superscriptaddress]{revtex4-1}

\usepackage{amsfonts}
\usepackage{amsmath}
\usepackage{amssymb}
\usepackage{bm}
\usepackage{dcolumn}
\usepackage{graphicx}
\usepackage{graphics}
\usepackage[latin1]{inputenc}
\usepackage{latexsym}
\usepackage{rotating}
\usepackage{hyperref}
\usepackage[all]{hypcap} 
\usepackage{xspace} 
\usepackage[usenames,dvipsnames]{color}
\usepackage{mathrsfs}
\usepackage{subfigure}
\usepackage{aas_macros}
\usepackage{amsfonts}
\usepackage{booktabs}
\usepackage{siunitx}
\usepackage{appendix}

\usepackage{ulem}
\usepackage{outlines}
\usepackage{enumitem}
\setenumerate[1]{label=\Roman*.}
\setenumerate[2]{label=\Alph*.}
\setenumerate[3]{label=\roman*.}
\setenumerate[4]{label=\alph*.}

\definecolor {darkgreen}{rgb}{0.2,0.7,0.2}

\newcommand\be{\begin{equation}}
\newcommand\ba{\begin{eqnarray}}
\newcommand\ee{\end{equation}}
\newcommand\ea{\end{eqnarray}}

\newcommand\bw{\begin{widetext}}
\newcommand\ew{\end{widetext}}

\newcommand{\GR}{{\mbox{\tiny GR}}}

\newcommand{\BH}{{\mbox{\tiny BH}}}
\newcommand{\EdGB}{{\mbox{\tiny EdGB}}}
\newcommand{\BD}{{\mbox{\tiny BD}}}
\newcommand{\ThreePN}{{\mbox{\tiny 3PN}}}

\newcommand{\NS}{*}

\begin{document}
\title{Constraining Gravity with Eccentric Gravitational Waves: \\ Projected Upper Bounds and Model Selection}

\author{Blake Moore}
\affiliation{eXtreme Gravity Institute, Department of Physics, Montana State University, Bozeman, MT 59717, USA.}

\author{Nicol\'as Yunes}
\affiliation{Department of Physics, University of Illinois at Urbana-Champaign, Urbana, IL 61801, USA.}

\date{\today}

\begin{abstract} 

Gravitational waves allow us to test general relativity in the highly dynamical regime. 
While current observations have been consistent with waves emitted by quasi-circular binaries, eccentric binaries may also produce detectable signals in the near future with ground- and space-based detectors.
We here explore how tests of general relativity scale with the orbital eccentricity of the source during the inspiral of compact objects up to $e \sim 0.8$.
We use a new, 3rd post-Newtonian-accurate, eccentric waveform model for the inspiral of compact objects, which is fast enough for Bayesian parameter estimation and model selection, and highly accurate for modeling moderately eccentric inspirals.
We derive and incorporate the eccentric corrections to this model induced in Brans-Dicke theory and in Einstein-dilaton-Gauss-Bonnet gravity at leading post-Newtonian order, which suggest a straightforward eccentric extension of the parameterized post-Einsteinian formalism. 
We explore the upper limits that could be set on the coupling parameters of these modified theories through both a confidence-interval- and Bayes-factor-based approach, using a Markov-Chain Monte Carlo and a trans-dimensional, reversible-jump, Markov-Chain Monte Carlo method.
We find projected constraints with signals from sources with $e \sim 0.4$ that are one order of magnitude stronger than that those obtained with quasi-circular binaries in advanced LIGO.
In particular, eccentric gravitational waves detected at design sensitivity should be able to constrain the Brans-Dicke coupling parameter $\omega \gtrsim  3300$ and the Gauss-Bonnet coupling parameter  $\alpha^{1/2} \lesssim 0.5 \; {\rm{km}}$ at 90\% confidence.
Although the projected constraint on $\omega$ is weaker than other current constraints, the projected constraint on $\alpha^{1/2}$ is 10 times stronger than the current gravitational wave bound.
\end{abstract}



\maketitle

\section{Introduction}
\label{intro}



The Laser Interferometer Gravitational-Wave Observatory (LIGO), its Italian counter-part Virgo, and other soon to be operational ground based gravitational wave (GW) detectors are positioned to probe the validity of general relativity (GR) and modified theories of gravity in the dynamical and non-linear strong field regime \cite{lrr-2013-9, will-living}. Indeed with the growing number of current detections \cite{2018arXiv181112907T}, we have been able to constrain the mass of the graviton, generic parameterized post-Einsteinian (ppE) deviations in the waveform, and the number spacetime dimensions to name a few effects \cite{2016PhRvL.116v1101A, 2019PhRvD.100j4036A, 2019PhRvL.123a1102A}. As we add new types of detectors and observe different types of sources, our ability to constrain modified gravity will only improve, as the strength of constraints depends heavily on the system observed. 

A particularly powerful source to test GR would be eccentric compact binary inspirals, but can we observe these signals with ground-based detectors?  All current observations are consistent with GWs produced by binaries in quasi-circular orbits \cite{2019arXiv190905466R, 2019arXiv190709384L}. However, there are a number of astrophysical channels by which we expect a small number of eccentric observations. Most of these scenarios rely on many-body interactions, such as scattering or the Kozai-Lidov mechanism in dense stellar regions. The most promising dense stellar environments for eccentric binary formation are globular clusters and regions near supermassive black holes. Rough estimates show that from globular clusters we could expect around $\sim 0.15$ eccentric events per Gpc per year with advanced LIGO (aLIGO) at design sensitivity\cite{2018PhRvL.120o1101R, 2018PhRvD..97j3014S, 2014ApJ...784...71S, 2019ApJ...871...91Z}. For the events around SMBH, about 0.5$\%$ are expected to have eccentricities greater than 0.1 when emitting GWs detectable by ground based detectors \cite{2012ApJ...757...27A, 2009MNRAS.395.2127O, 2014ApJ...781...45A}. 

Even if one were able to observe GWs from eccentric compact binary inspirals, one may still wonder precisely what one gains from observing these signals as compared to observing quasi-circular inspirals. GWs from eccentric inspirals contain amplitude and phase modulations induced in part by precession that are typically absent in their quasi-circular counterparts. These modulations contain information that, if extracted properly, can lead to a more accurate and robust estimation of the parameters of the system. For example, eccentric signals can provide a factor of 1-2 improvement in sky localization \cite{2019arXiv191204455P, 2017PhRvD..96h4046M}, and factors of 10 or more increase in accuracy of measurement of the source parameters \cite{Gond_n_2018, mydata}. Naturally, following from the argument that eccentric binaries are formed in specific stellar regions, an ensemble of eccentric detections can provide constraints on formation scenarios with as little as tens of observations \cite{2019MNRAS.486..570T, 2016PhRvD..94f4020N, 2017MNRAS.465.4375N}. 

Eccentric signals could be an optimal source for validating GR and constraining alternative theories of gravity given the promising results in terms of sky localization and parameter measurement. References \cite{2013IJMPD..2241013Y, 2010PhRvD..81f4008Y} studied how projected constraints on several modified theories of gravity depend on eccentricity using space-based detectors. These papers, however, only considered weakly-eccentric (and weakly-spinning) compact binary inspirals, and thus, they ignored eccentric corrections to non-GR terms in the waveform model. This was because if the non-GR effect is assumed to be small, and if the eccentricity is assumed to be small, then eccentric corrections to non-GR terms are a second order effect, which was ignored in those papers.   Reference \cite{2019arXiv190807089M} lifted this later assumption, calculating and incorporating eccentric corrections in the non-GR sector of the waveform explicitly. This paper confirmed that, at low eccentricity, covariances between non-GR and GR parameters in the waveform model deteriorate our ability to constrain GR; nonetheless, when the eccentricity of the signal is large enough, then these covariances begin to break, suggesting the constraints on non-GR effects could recover and become stronger than quasi-circular constraints for sufficiently eccentric observations.  The study of Ref.~\cite{2019arXiv190807089M}, however, was limited to mildly eccentric inspirals because the waveform model used became highly inaccurate of eccentricities larger than $e \sim 0.1$--$0.2$.      

From the above summary, it should be clear that all previous studies have been limited to the small eccentricity regime because no model existed that was shown to be valid to high eccentricities, while remaining fast enough to carry out parameter estimation studies. In Ref.~\cite{my3PN}, we recently developed a 3 post-Newtonian (PN\footnote{A PN waveform is one constructed assuming small velocities and weak fields. In particular, a 3PN waveform is one which contains relativistic corrections of relative ${\cal{O}}(v^{6}/c^{6})$ with respect to the controlling factor in the expansion~\cite{Blanchet:2002av}.}) accurate waveform, TaylorF2e, to model the eccentric inspiral of compact binaries, which has been validated through a parameter bias and overlap study for eccentricities as high as $\sim 0.8$. Moreover, this waveform model is fast enough for use in parameter estimation studies, which employ Markov Chain Monte Carlo (MCMC) methods \cite{mydata}. This model, therefore, provides the foundations on which to build non-GR eccentric waveforms and study how well we can test GR with moderately eccentric signals. 

In this paper, we begin by deriving non-GR corrections to TaylorF2e due to leading PN order effects in (massless) Brans-Dicke (BD) and in Einstein-dilaton-Gauss-Bonnet (EdGB) theory. In BD theory, a dynamical scalar field couples to the physical (Jordan frame) metric to modify how massive (strongly self-gravitating) bodies move, thus violating the strong equivalence principle~\cite{lrr-2013-9}. In EdGB theory\footnote{EdGB gravity defines a class of theories in which a dynamical scalar field couples non-minimally to the Gauss-Bonnet invariant in the action through a coupling function. We here study EdGB gravity with a linear coupling function, which is sometimes also referred to as scalar-Gauss-Bonnet gravity.}, a dynamical scalar field couples to the curvature, also introducing violations to the strong equivalence principle~\cite{lrr-2013-9}. In both of these theories, the dynamical scalar field introduces a -1PN modification to the rate of change of the orbital energy and angular momentum, which in turn changes the orbital dynamics and the GWs emitted. 

With these corrections derived, we turn to analyzing the ability of a single aLIGO detector (at design sensitivity) to constrain either theory. This is explored with two different techniques. One technique is to synthesize a GR signal injection and to attempt to recover it with a non-GR waveform template or model through an MCMC exploration of the likelihood function in Bayesian parameter estimation. After such an exploration, we can estimate an upper limit on the non-GR coupling parameter by explicitly calculating the confidence intervals from its marginalized posterior distribution. This is similar to a Fisher matrix analysis, except that the latter one assumes a Gaussian posterior and uses the standard deviation of that posterior to estimate confidence intervals. Another approach is to carry out a Bayes factor analysis to investigate at which value of the coupling parameter the effects of the alternative theory are significant enough such that the data supports the alternative theory over GR. Since such a study can be heavily influenced by the priors, we also explore how different priors can affect our results.

This work has led to two main results. First, in Eqs.\eqref{eq:phase_func} and \eqref{eq:y_alph_eexp} we provide the analytic expressions that are necessary to construct the Fourier transform of the GW response function in the TaylorF2e model for both BD theory and EdGB gravity. Interestingly, we find that the corrections due to both theories can be very simply mapped to one another through a relation between their coupling parameters. This suggests that a general formalism for eccentric waveforms where different theories can be mapped to the same set of corrections, like in the ppE formalism in \cite{PPE}, must exist. We also find that although previous studies found deteriorated constraints on alternative theories of gravity when including eccentricity, this is likely due to only working in the small eccentricity regime, as Ref.~\cite{2019arXiv190807089M} hinted at. At higher eccentricities, the strong covariances between parameters, which sourced the deteriorated constraint, are broken, and the ability to constrain EdGB and BD increases by about one order of magnitude as the eccentricity of the signal is increased from 0 to $\sim 0.4$. 

\begin{figure}[htp]
\includegraphics[clip=true,angle=0,width=0.475\textwidth]{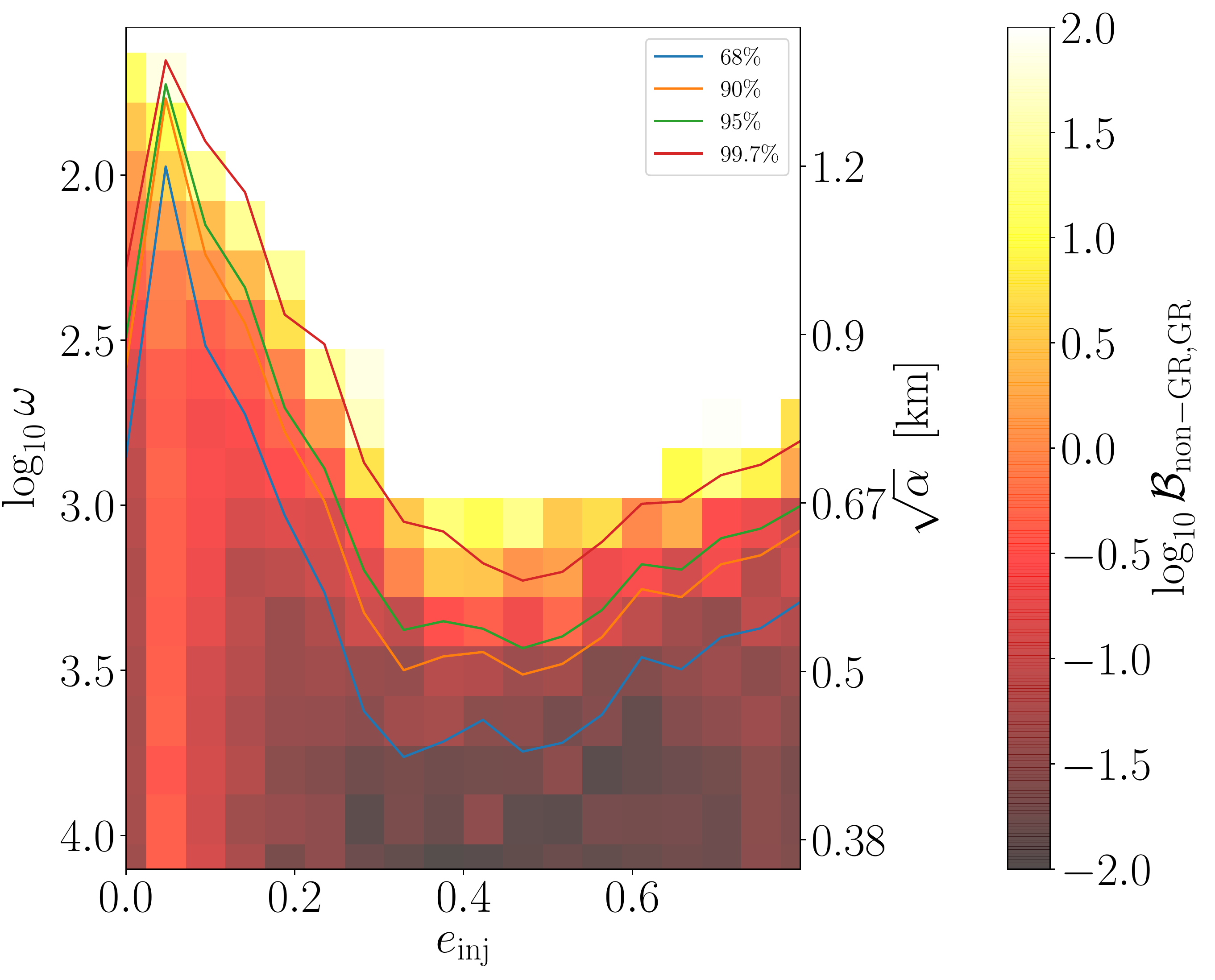}
\caption{\label{fig:intro_PE} A heat plot of the Bayes factor in favor of the non-GR model as a function of injected eccentricity and injected non-GR coupling parameter on the y-axis for a $(10, 1.4)M_{\odot}$ system with SNR 30. Values above 2 and below -2 have been excluded. We have overplotted the upper bounds on the non-GR parameters obtained through requiring different confidence regions as indicated in the legend. In both the upper bounds from the confidence interval analysis and Bayes Factor analysis we see nearly an order of magnitude increased constraint on GR in Brans Dicke ($\omega$) and about a factor of 2 in EdGB ($\sqrt{\alpha}$). }
\end{figure}
Figure \ref{fig:intro_PE} shows the projected (log-10)Bayes factor as a function of the injected coupling parameter of BD theory $\omega$ and EdGB gravity $\alpha^{1/2}$, and as a function of the orbital eccentricity of the signal for a source with component masses $(10, 1.4) M_{\odot}$ and SNR 30. This figure also shows upper bounds (confidence intervals) on the coupling parameters obtained through a Bayesian parameter estimation study. Observe that confidence limit constraints are consistent with the Bayes factor constraint, with a $3\sigma$ bound ($99.7\%$ confidence interval) corresponding roughly to a Bayes factor of $10$. Observe also that our analysis suggests the best constraints come from the inspiral of compact binaries with orbital eccentricities of $e \sim 0.4$, leading to constraints on $\alpha^{1/2} \lesssim 0.6$ km and $\omega \gtrsim 10^{3}$ at $3\sigma$; due to historical reasons, the GR limit is recovered when $\omega \rightarrow \infty$. This deterioration of the constraint past eccentricities of $0.4$ is consistent with the number of phase cycles in the signal. The latter initially increases as the number of harmonics of mean motion in the signal increases, because each harmonic contributes its own number of phase cycles. As the eccentricity is increased past $\sim 0.4$, the number of phase cycles begins to decrease because eccentric binaries inspiral faster.

How do these projected constraints compare to other current bounds on EdGB gravity and BD theory? The most robust and most stringent current bound on the EdGB coupling parameter $\alpha^{1/2}$ 
comes from a recent analysis of GW170608, which set a $90\%$ bound of 
$\alpha^{1/2} \lesssim 5.6 \; {\rm{km}}$~\cite{2019PhRvL.123s1101N}. 
This is to be compared with a projected $90\%$ bound in Fig.~\ref{fig:intro_PE}, namely $\alpha^{1/2} \lesssim  0.5$ km for a signal with $e \sim 0.4$, which is about one order of magnitude stronger. On the other hand, the most stringent constraint currently placed on the BD coupling parameter $\omega$ comes from measuring the frequency shift of radio photons as they passed near the Sun to and from the Cassini spacecraft, which sets $\omega \gtrsim 40,000$ \cite{Bertotti:2003rm} at $95\%$ confidence. Our projected constraints in Figure \ref{fig:intro_PE} are not a strong as the Cassini bound ($\omega \gtrsim 2700$) at 95\% confidence, even when the GW constraint is boosted by the effects of orbital eccentricity in the signal. 

The paper is organized as follows. In Section \ref{sec:phasing} we derive the EdGB and BD eccentric corrections to the orbital dynamics and frequency response. Section \ref{sec:bayes_frame} presents the Bayesian framework for our analysis, the notation used, and derives some useful approximations. The results of our model selection and confidence interval based approaches to exploring the ability of eccentric signals to constrain EdGB and BD are presented in Section \ref{sec:results}. Lastly we point to interesting future extensions suggested by this work and general conclusions in Section \ref{sec:conc}. Throughout we have set $c = G = 1$.
\section{Modified Gravity Corrections to Eccentric Signals}
\label{sec:phasing}
In this section, we present the incorporation and derivation of the effects of EdGB and BD gravity into our existing 3PN GR eccentric waveform. We will not provide here a detailed description of BD or EdGB theory, but instead refer the interested reader to~\cite{lrr-2013-9}. Each of these theories introduce a dynamical scalar field in the Einstein-Hilbert action, which in BD is sourced by the matter stress energy tensor, while in EdGB by a curvature-squared invariant. In turn, these theories both admit dipole radiation, which adds corrections to the energy and angular momentum flux of the binary, and therefore, changes to the gravitational waveform. We work to leading PN order in the non-GR corrections while retaining the 3PN accuracy in the GR sector of our eccentric model derived in \cite{my3PN} and optimized in \cite{mydata}. We find here that because the non-GR corrections take a similar form at this order it is simple to map the corrections induced by the two theories to one another.

In order to obtain the eccentric non-GR corrections to the frequency response, we follow the prescription laid out in \cite{my3PN} and optimizations laid out in \cite{mydata}. The 3PN valid eccentric model schematically takes the form: 
\begin{align}
\label{eq:spa_simp}
\tilde{h}(f) &= \mathcal{A}\sum_{j=-1}^{15}N_j\sqrt{\frac{y^{-7}}{(j+2)(96 + 292e^2 + 37e^4)}}e^{i\psi_j} \,,
\end{align}
where the $j^{\rm th}$ Fourier phase is given by
\begin{equation}
\label{eq:phases}
\psi_j = 2\pi f t(e) - j l(e) - 2\lambda(e) - \frac{\pi}{4}\,. 
\end{equation} 
In the above, $\mathcal{A}$ is an overall amplitude term which depends on the masses, distance to source, and source orientation. The $N_j$ are harmonic amplitudes which control the contribution of each harmonic and scale as $e^{|j|}$. Our post-Newtonian expansion parameter $y$ is related to the semi-latus rectum ($p$) and orbital frequency ($\omega$) by:
\begin{equation}
y = \sqrt{\frac{M}{p}} = \frac{(M\omega)^{1/3}}{\sqrt{1-e^2}} \, .
\end{equation}
Here and above we work with the time-eccentricity $e_{t} = e$, $\omega$ is the azimuthal orbital frequency and $M$ is the total mass of the system. To specify each of the Fourier phases we must model the different phase functions appearing in Eq.\eqref{eq:phases}: the time to coalescence $t(e)$, the mean anomaly $l(e)$, and $\lambda(e)$, which is an orbital angle related to the azimuthal orbital frequency $\omega$. Since, at 1PN order, eccentric orbits undergo periastron precession, there are two comparable orbital scales. One is related to the azimuthal period ($P_{\phi}$), and the other is related to the radial period, related to the periastron to periastron period ($P_{r}$). These two differing scales give rise to the orbital angles $\lambda(e)$ and $l(e)$ respectively, and the orbital frequencies $\omega$ and $n$.   

Our strategy is to specify all of the functions appearing in Eq.~\eqref{eq:spa_simp} analytically in terms of the eccentricity. This can succinctly be seen as solving the following set of equations:
\begin{subequations}
\label{eq:sys_eq}
\begin{align}
y(e) &= \int^{e}\frac{dy}{de'}de' \, ,  \label{eq:sys_eq_y} \\
t(e) &= \int^{e}\frac{dt}{de'}de' \, , \label{eq:sys_eq_t}\\
\lambda(e) &= \int^{e} \omega(e')\frac{dt}{de'}de' \, ,  \label{eq:sys_eq_lam}\\
l(e) &= \int^{e} n(e')\frac{dt}{de'}de' \, .  \label{eq:sys_eq_l}
\end{align}
\end{subequations}
In the above, the two orbital frequencies $n$ and $\omega$ as well as $dt/de$ are originally functions of both $y$ and $e$, so first $y(e)$ must be specified (i.e.~Eq.~\eqref{eq:sys_eq_y} must be solved first). Then, $y(e)$ is substituted into the integrands of the phase functions in Eqs.~\eqref{eq:sys_eq_t}-\eqref{eq:sys_eq_l}, and the integrands are expanded in eccentricity and finally integrated. We then map Fourier frequency to the eccentricity using the stationary phase condition. For much more detail in deriving these corrections and computing the Fourier response, see \cite{my3PN}.

In order to solve Eqs.~\eqref{eq:sys_eq_y}-\eqref{eq:sys_eq_l} with the non-GR corrections, we introduce a small parameter $\kappa$ that vanishes in the GR limit and takes a small, but non-zero value in the non-GR theory. We then expand to leading order in $\kappa$ and to leading PN order in the corrections that are proportional to $\kappa$. The theories we work with here are already constrained to the level that higher PN order corrections will not change the conclusions of our work (see e.g.~Appendices A and B in~\cite{Yunes:2016jcc} for the effect of higher PN order terms in inspiral waveforms for tests of GR). The integrands in Eqs.~\eqref{eq:sys_eq_y}-\eqref{eq:sys_eq_l} are then given as a series in $\kappa$ and $e$. This will then provide a set of non-GR corrections that are linearly proportional to $\kappa$: $y_{\kappa}(e)$, $t_{\kappa}(e)$, $l_{\kappa}(e)$, and $\lambda_{\kappa}(e)$, from which we can specify the full function to be used in the frequency response as $y(e) = y_{\ThreePN,\GR}(e) + y_{\kappa}(e)$ and so forth for the other necessary functions. We use the 3PN accurate functions for the GR corrections given in \cite{my3PN}. In Sec.~\ref{subsec:orb_dyn} we compute in more detail the non-GR corrections to $\dot{e}$ and $\dot{y}$ required to form the integrands in Eqs.~\eqref{eq:sys_eq_y}-\eqref{eq:sys_eq_l}. In Sec.~\ref{subsec:freq_resp} we solve Eqs.~\eqref{eq:sys_eq_y}-\eqref{eq:sys_eq_l} and give the explicit functional forms of $y_{\kappa}(e)$, $t_{\kappa}(e)$, $l_{\kappa}(e)$, and $\lambda_{\kappa}(e)$ that are used in our analysis.

\subsection{Orbital Dynamics}
\label{subsec:orb_dyn}
Here we derive the differential equations $\dot{y}$ and $\dot{e}$ that are used to express the integrands in Eqs.~\eqref{eq:sys_eq_y}-\eqref{eq:sys_eq_l} in terms of only $e$ and constants in both EdGB and BD. In order to derive these equations we start with the expressions for the binding energy, $E$, and the angular momentum, $L$, of the binary and their time derivatives. 

In BD theory, the binding energy and (the $\hat{z}$ component of the) angular momentum at leading PN order are left unchanged from GR, and are thus given by:
\begin{subequations}
\label{eq:e_l_gr}
\begin{align}
E_{\GR} &= \frac{M\eta}{2}(e^2 - 1)y^2 \, , \\
L_{\GR} &= \frac{M^2 \eta}{y} \,.
\end{align}
\end{subequations}
The rate of change of these two quantities, however, is modified in BD theory, and in our notation they take the form $\dot{E} = \dot{E}_{\GR} + \dot{E}_{\BD}$ and $\dot{L} = \dot{L}_{\GR} + \dot{L}_{\BD}$, where the GR fluxes are\cite{2017CQGra..34m5011L}
\begin{subequations}
\label{eq:e_l_dot_GR}
\begin{align}
\dot{E}_{\GR} &= -\frac{1}{15} \eta^2 y^{10}(1-e^2)^{3/2}(96 + 292e^2 + 37e^4) \, , \\
\dot{L}_{\GR} &= -\frac{4}{5} M\eta^2 y^7(1 - e^2)^{3/2}(8 + 7e^2)  \, , \\
\end{align}
\end{subequations}
and the BD corrections are
\begin{subequations}
\label{eq:e_l_dot_bd}
\begin{align}
\dot{E}_{\BD} &= - \frac{32}{5} \eta^2 b y^8(1-e^2)^{3/2} \left(1 + \frac{e^2}{2} \right) \, , \\
\dot{L}_{\BD} &=  -\frac{32}{5} M\eta^2 b y^5(1 - e^2)^{3/2} \, .
\end{align}
\end{subequations}
to leading order in $\kappa$ and to leading PN order. The BD corrections here follow the conventions of \cite{Will:1989sk}, where $b = 5 \mathcal{S}_{\BD}^2/48\omega_{\BD}$, $\mathcal{S}_{\BD}$ is the sensitivity difference of the two objects ($\mathcal{S}_{\BD} = s_1 - s_2$), and $\omega_{\BD}$ is the BD coupling parameter, which goes to infinity in the GR limit. In this theory, black holes have $s_{\BH} = 0.5$, and neutron stars have sensitivities of $s_{\NS} \sim 0.15$, depending on their equation of state. 


With this at hand, we can now compute $\dot{y}$ and $\dot{e}$ in terms of $y$ and $e$. To do so, we use implicit differentiation, $\dot{y}$ and $\dot{e}$ in terms of $\dot{E}$ and $\dot{L}$ through Eqs.~\eqref{eq:e_l_gr}, then substitute $\dot{E}$ and $\dot{L}$ with Eqs.~\eqref{eq:e_l_dot_bd}. The result is then expanded to leading order in a $b \ll 1$ expansion. We are also considering the GR terms up to 3PN order for our later analysis, but for readability and simplicity of deriving the non-GR terms, we will here omit the GR higher order PN terms here. We then have $\dot{y} = \dot{y}_{\GR} + \dot{y}_{\BD}$ and $\dot{e} = \dot{e}_{\GR} + \dot{e}_{\BD}$, where the GR terms are
\begin{subequations}
\label{eq:edot_ydot_GR}
\begin{align}
\dot{y}_{\GR} &= \frac{4}{5} \frac{\eta}{M}y^9(1-e^2)^{3/2}(8+7e^2) \, , \label{eq:ydot_gr} \\
\dot{e}_{\GR} &= - \frac{1}{15} \frac{\eta}{M} y^8e(1-e^2)^{3/2}(304+121e^2) \, , \label{eq:edot_gr} 
\end{align}
\end{subequations}
and the BD corrections re
\begin{subequations}
\label{eq:edot_ydot_BD}
\begin{align}
\dot{y}_{\BD} &= \frac{32}{5} \frac{\eta}{M}b y^7(1-e^2)^{3/2} \, , \\
\dot{e}_{\BD} &= - \frac{48}{5} \frac{\eta}{M}b y^6e(1-e^2)^{3/2} \, .
\end{align}
\end{subequations}
The BD corrections enter at -1PN order relative to the GR terms, which suggests their effect may be large enough to be constrainable. With this in hand, we are now in a position to solve Eqs.~\eqref{eq:sys_eq} to prescribe the Fourier response, but let us first derive similar corrections in the orbital dynamics for EdGB gravity.

As in BD theory, the binding energy and (the $\hat{z}$ component of the) angular momentum in EdGB gravity are not modified to leading PN order \cite{2017CQGra..34m5011L}. This then means that these quantities can be approximated by their GR expressions, which at leading PN order were given in Eq.~\eqref{eq:e_l_gr}.
The fluxes of energy and angular momentum, however, are modified in EdGB gravity to $\dot{E} = \dot{E}_{\GR} + \dot{E}_{\EdGB}$ and $\dot{L} = \dot{L}_{\GR} + \dot{L}_{\EdGB}$~
\cite{2017CQGra..34m5011L}, where $\dot{E}_{\GR}$ and $\dot{L}_{\GR}$ are given in Eqs.~\eqref{eq:e_l_dot_GR}, while 
\begin{subequations}
\label{eq:e_l_dot_edgb}
\begin{align}
\dot{E}_{\EdGB} &= -\frac{\eta^2}{3}\mathcal{S}_{\EdGB}^2 y^8(1-e^2)^{3/2}  \left(1+\frac{e^2}{2}\right) \, , \\ 
\dot{L}_{\EdGB} &= -\frac{M\eta^2}{3}\mathcal{S}_{\EdGB}^2 y^5 (1-e^2)^{3/2} \, ,
\end{align}
\end{subequations}
where $\mathcal{S}_{\EdGB} = \zeta_{1}^{1/2} - \zeta_{2}^{1/2}$ and $\zeta_{1,2} = \xi/m_{1,2}^4$ are dimensionless deformation parameters, with $\xi = 16 \pi \alpha^{2}$ and $\alpha$ the EdGB coupling parameter with units of length squared. If one of the objects is a neutron star, its associated $\zeta_{n}$ vanishes, and modifications to the fluxes of energy and angular momentum enter at higher PN order.  

As in the BD theory case, we can use implicit differentiation to solve the system of equations for $\dot{y}_{\EdGB}$ and $\dot{e}_{\EdGB}$. The GR corrections were given in Eqs.~\eqref{eq:ydot_gr}-\eqref{eq:edot_gr} already, so here we only list the corrections due EdGB gravity:
\begin{subequations}
\label{eq:edot_ydot_EDGB}
\begin{align}
\dot{y}_{\EdGB} &= \mathcal{S}_{\EdGB}^2\frac{\eta}{3M}y^7(1-e^2)^{3/2}  \, , \\
\dot{e}_{\EdGB} &= -\mathcal{S}_{\EdGB}^2\frac{\eta}{2M}y^6 e (1-e^2)^{3/2} \, .
\end{align}
\end{subequations}
By inspection of Eqs.~\eqref{eq:edot_ydot_BD} and~\eqref{eq:edot_ydot_EDGB} we realize that there is a simple mapping between the equations $\dot{y}$ and $\dot{e}$ between the two theories: $96b = 5\mathcal{S}_{\EdGB}^2$. This suggests that we ought to parameterize $\dot{y}$ and $\dot{e}$ in terms of a small parameter $\kappa$ which can be mapped to either theory via $\kappa = 96b$ to map to BD theory or $\kappa =  5\mathcal{S}_{\EdGB}^2$ to map to EdGB provided
\begin{subequations}
\begin{align}
\label{eq:edot_ydot_alpha}
\dot{y}_{\kappa} &= \kappa\frac{\eta}{15M}y^7(1-e^2)^{3/2}  \, , \\
\dot{e}_{\kappa} &= -\kappa\frac{\eta}{10M}y^6 e (1-e^2)^{3/2}\, .
\end{align}
\end{subequations}
Moving forward we present phasing results in terms of $\kappa$, as it is very simple to convert the results to either of the theories through the above mapping.

In what regime in $\kappa$ are we justified in treating the corrections introduced by the modified theories as a small deformation of the PN dynamics? This region is defined by requiring that the non-GR corrections to the orbital dynamics ($\dot{y}, \dot{e}$) be smaller than the GR PN dynamics. The most conservative estimate of where our expansions in $\kappa$ are valid can be found from
\begin{equation}
\dot{y} = \frac{32}{5}\frac{\eta}{M}\left(y^9 + \kappa\frac{y^7}{96}\right) \, ,
\end{equation}
where we have assumed circularity for simplicity. Inspection of the above equation suggests that our expansion is valid provided $\kappa \ll 96y^2$. Assuming circularity again, we can define the region in which the expansion in $\kappa$ is valid to be
\begin{equation}
\label{eq:alp_constraint}
\kappa \ll 0.06 ~ \left(\frac{M}{M_{\odot}}\right)^{2/3} \left(\frac{f_{\rm low}}{1 \; {\rm{Hz}}}\right)^{2/3}\,,
\end{equation}
where in the above $f_{\rm low}$ is the lower frequency cutoff of the detector. For a $(10, 10)M_{\odot}$ or $(1.4, 1.4)M_{\odot}$ system with a detector lower frequency cutoff of 10Hz, this translates to $\kappa \ll 2.04$ and $0.550$ respectively, while for a system with $M = 10^6 M_{\odot}$ and a detector with $f_{\rm low} = 10^{-5}$Hz, Eq.~\eqref{eq:alp_constraint} implies $\kappa \ll 0.276$. If we take the more conservative of the bounds for the aLIGO-like sources, we can then set $\omega \gg 0.182$ (assuming $\mathcal{S}_{\rm BD} \sim \mathcal{O}(10^{-1})$). Current constraints place us well within this small $\kappa$ limit, and our region of interest for our Bayesian analysis will lie within this limit.

The mapping between $\kappa$ and $\alpha$ depends strongly on the mass difference and whether one of the objects is a neutron star. In this paper we explore an optimal system with component masses $(1.4, 10)M_{\odot}$. The expansions in $\kappa$ are then valid when $\alpha^{1/2} \ll 4$ km when we assume $f_{\rm low} = 10$ Hz and that the NS is not (monopolarly) charged under the scalar field (as is the case in EdGB gravity). Reference \cite{2019PhRvL.123s1101N} used GW170608 to place a 90\% constraint on EdGB of $\alpha^{1/2} < 5.6$ km using the GWs emitted by quasi-circular binary; for such a binary, this constraint is at the edge of the region of validity of the small coupling approximation. Do note, however, that $\alpha^{1/2}$ scales as $\kappa^{1/4}$ so an order of magnitude difference in $\alpha^{1/2}$ scales as 2 orders of magnitude in $\kappa$. Thus, we could be working just below these limits and still be well within the limit on $\kappa$.  In our Bayesian analysis we place an upper limit on the prior on $\kappa$ which ensures we are working well within the limit where our expansions in $\kappa$ can be treated as a deformation of the PN GR dynamics.

Let us now make some general comments regarding the above modified gravity modifications to the rate of change of the PN parameter $y$ and the eccentricity $e$. Since both $y$ and $e$ are dimensionless, on general grounds we would expect $\dot{e}_{\kappa}$ and $\dot{y}_{\kappa}$ to be proportional to $M^{-1}$. Moreover, since we assume the deformation from GR possesses a continuous GR limit, and the deformation is small (in the sense that we can linearize about the GR background), then both of these rates of change must be linear in $\kappa$. Furthermore, since the modifications to the energy and angular momentum fluxes we are considering are both of -1PN order relative to GR, we then expect $\dot{y}_{\kappa} \sim y^{7}$ and $\dot{e}_{\kappa} \sim y^{6}$. Whether general principles can be used to deduce the $\eta$, and especially the $e$ dependence of $\dot{y}_{\kappa}$ and $\dot{e}_{\kappa}$ requires further study. Either way, the above analysis suggests that a ppE generalization of the TaylorF2e model is possible. 

\subsection{Frequency Response}
\label{subsec:freq_resp}
Let us proceed by using the expressions for $\dot{y}_{\kappa}$ and $\dot{e}_{\kappa}$ to produce the non-GR corrections to the functions required to compute the frequency response set forth in Eqs.~\eqref{eq:sys_eq}: $y_{\kappa}$, $t_{\kappa}$, $l_{\kappa}$, $\lambda_{\kappa}$. For compactness we will often not give the GR expressions for these functions as they can be found in \cite{my3PN} and \cite{2018CQGra..35w5006M}. Let us first derive an analytic solution for the frequency evolution as a function of eccentricity, $y(e)$. This is obtained by direct integration of $dy/de = \dot{y}/\dot{e}$, employing expansions in $\kappa \ll 1$ throughout.

Expanding $dy/de$ to leading order in $\kappa$ and integrating yields an exact solution:
\begin{widetext}
\begin{equation}
\label{eq:ye_exact}
y(e) = \sqrt{\frac{C_{1}}{e^{12/19}(304+121e^2)^{870/2299}}-\kappa\left[\frac{396720}{1429(304+121e^2)}-\frac{3230{}_2F_1(\frac{84}{121},1;\frac{25}{19};-\frac{121}{304}e^2)}{4287}\right]} \, ,
\end{equation}
where ${}_2F_1(a,b;c;z)$ is the ordinary hypergeometric function. The constant of integration, $C_1$, is determined by requiring that $y(e_0) = y_0$:
\begin{equation}
\label{eq:C1_exact}
C_1 = e_0^{12/19}(304+121e_0^2)^{870/2299}\left\lbrace y_0^2-\kappa\left[\frac{396720}{1429(304+121e_0^2)}-\frac{3230{}_2F_1(\frac{84}{121},1;\frac{25}{19};-\frac{121}{304}e_0^2)}{4287}\right]\right\rbrace \, .
\end{equation}
After plugging in this constant into Eq.~\eqref{eq:ye_exact} and expanding to leading order in $\kappa$ we have 
\begin{align}
\label{eq:ye_1stexp}
y_{\GR}(e) &= \frac{e_0^{6/19} \left(121 e_0^2+304\right){}^{435/2299} y_0}{e^{6/19} \left(121 e^2+304\right)^{435/2299}} \\
y_{\kappa}(e) &= 5 \kappa  \left\lbrace5168 \left(\frac{121 e^2}{304}+1\right)^{1429/2299} e_0^{12/19} \left(\frac{121 e_0^2}{304}+1\right) \,
   _2F_1\left(\frac{84}{121},1;\frac{25}{19};-\frac{121 e_0^2}{304}\right) 
   \right.  \nonumber \\ & \left. 
  -5168 e^{12/19} \left(\frac{121 e^2}{304}+1\right) \left(\frac{121
   e_0^2}{304}+1\right){}^{1429/2299} \, _2F_1\left(\frac{84}{121},1;\frac{25}{19};-\frac{121 e^2}{304}\right)
   \right. \nonumber \\ & \left.
   +6264 \left[\left(\frac{121
   e^2}{304}+1\right)^{1429/2299} e_0^{12/19}-e^{12/19} \left(\frac{121 e_0^2}{304}+1\right){}^{1429/2299}\right]\right\rbrace
   \nonumber \\ &
   \times \left[20851968 e^{6/19}
   \left(\frac{121 e^2}{304}+1\right)^{1864/2299} e_0^{6/19} \left(\frac{121 e_0^2}{304}+1\right){}^{1864/2299} y_0\right]^{-1}
\end{align}
While the above equation is exact, the sampling of the hypergeometric functions can become costly and we found in \cite{2018CQGra..35w5006M} that these special functions could very faithfully be represented, even to high eccentricities, in a low eccentricity expansion without keeping a computationally prohibitive number of terms. To be consistent, we choose to also expand all other terms in eccentricity as well. What results for the expanded $\kappa$ correction to $y$ is the sum of two Taylor series with different controlling factors,
\begin{align}
\label{eq:y_alph_eexp}
y_{\kappa}(e) &= -\frac{5 \kappa  e^{6/19}}{1824 e_0^{6/19} \left(\frac{121 e_0^2}{304}+1\right){}^{435/2299} y_0}
\left(1-\frac{17163 e^2}{72200}+\frac{72226161 e^4}{917459840}-\frac{447503490791 e^6}{15897744107520}+\frac{79594709788763
   e^8}{7652114163752960}
	\right. \nonumber \\ & \left.   
   -\frac{69877537192796697791 e^{10}}{17856239009574187171840}\right) + \frac{5 \kappa  e_0^{6/19} \left[646 \left(\frac{121 e_0^2}{304}+1\right) \, _2F_1\left(\frac{84}{121},1;\frac{25}{19};-\frac{121
   e_0^2}{304}\right)+783\right]}{2606496 e^{6/19} \left(\frac{121 e_0^2}{304}+1\right){}^{1864/2299} y_0}
    \nonumber \\ & \times \left(1-\frac{435 e^2}{5776}+\frac{594645 e^4}{33362176}-\frac{997616095 e^6}{192699928576}+\frac{1828630302135
   e^8}{1113034787454976}-\frac{3522307687972437 e^{10}}{6428888932339941376}\right) \, .
\end{align}
\end{widetext}
We keep the hypergeometric function which is a function of the initial eccentricity $e_{0}$ unexpanded as this need only be evaluated once per entire waveform generation so its cost is negligible. Note that although the second term above scales as $e^{-6}$, it obviously does not diverge faster as $e \to 0$ than the GR term, which as we can see in Eq.~\eqref{eq:ye_1stexp} diverges at the same rate.

With $y(e)$ now in hand, we seek the phase functions $t(e)$, $l(e)$, and $\lambda(e)$ which were introduced in Eqs.~\eqref{eq:sys_eq}. For simplicity we will only list the non-GR contributions $t_{\kappa}(e)$, $l_{\kappa}(e)$, and $\lambda_{\kappa}(e)$. Since we are working at leading PN order in the non-GR terms and periastron precession is a 1PN effect we have then that
\begin{equation}
\label{eq:l_alpha}
l_{\kappa}(e) = \lambda_{\kappa}(e) = \int^{e} \left[\frac{y(e')^3(1-e'^2)^{3/2}}{M}\frac{dt}{de'}\right]_{\kappa}de' \, ,
\end{equation}
where $[~ ~ ~]_{\kappa}$ is shorthand notation for the expansion of the term within the brackets in $\kappa \ll 1$ and taking only the term which scales as $\kappa$ (as the term which scales as $\kappa^{0}$ is the GR term which has already been derived in the literature). Likewise for $t_{\kappa}(e)$ then
\begin{equation}
\label{eq:t_alpha}
t_{\kappa}(e) = \int^{e} \left[\frac{dt}{de'}\right]_{\kappa}de' \, ,
\end{equation}
where in Eqs.~\eqref{eq:l_alpha} and~\eqref{eq:t_alpha} it is understood that we have substituted Eq.~\eqref{eq:ye_1stexp} into the integrands in order to obtain them as functions of the eccentricity and constants only. Some of the terms which appear in these integrands can be integrated directly giving rise to AppellF1 functions and more $_{2}F_{1}$ hypergeometric functions; however, since we intend to series expand these expressions in $e \ll 1$, and since series expansion and integration commute in our case (due to the uniform nature of the expansions), it is generally simpler to expand the integrands in eccentricity first, and then integrate the result. These integrations yield a similar structure to that of Eq.~\eqref{eq:y_alph_eexp}:
\begin{widetext}
\begin{subequations}
\label{eq:phase_func}
\begin{align}
l_{\kappa}(e) &= -\frac{5 \kappa  e^{42/19}}{25536 e_0^{42/19} \left(\frac{121 e_0^2}{304}+1\right){}^{3045/2299} \eta  y_0^7}\left(1-\frac{132027 e^2}{1848320}+\frac{293611227 e^4}{21652052224}-\frac{2213979348413 e^6}{661346154872832}+\frac{2223905948814193
   e^8}{2375216236428918784}
   \right. \nonumber \\ & \left.
   -\frac{1867393272311822254223 e^{10}}{6628235920353938278187008}\right) + \frac{25 \kappa  e^{30/19} \left(646 \left(\frac{121 e_0^2}{304}+1\right) \, _2F_1\left(\frac{84}{121},1;\frac{25}{19};-\frac{121
   e_0^2}{304}\right)+783\right)}{83407872 e_0^{30/19} \left(\frac{121 e_0^2}{304}+1\right){}^{4474/2299} \eta  y_0^7}
   \nonumber \\ & \times
   \left(1-\frac{465 e^2}{49096}+\frac{1126695 e^4}{884097664}-\frac{295569655 e^6}{1156199571456}+\frac{889369091895
   e^8}{14469452236914688}-\frac{290112197776149 e^{10}}{17679444563934838784}\right)\,,  \\
t_{\kappa}(e) &= -\frac{31 \kappa  e^{60/19} M}{116736 e_0^{60/19} \left(\frac{121 e_0^2}{304}+1\right){}^{4350/2299} \eta  y_0^{10}}\left(1+\frac{526002 e^2}{548359}+\frac{83906652315 e^4}{96700267136}+\frac{25781287733225 e^6}{32481981710592}
 \right. \nonumber \\ & \left.
+\frac{238136880469635965
   e^8}{324449640543125504}+\frac{237106477791545365546321 e^{10}}{345964630810499876454400}\right) +
	\nonumber \\ &   
 \frac{25 \kappa  e^{48/19} \left(646 \left(\frac{121 e_0^2}{304}+1\right) \, _2F_1\left(\frac{84}{121},1;\frac{25}{19};-\frac{121
   e_0^2}{304}\right)+783\right)M}{83407872 e_0^{48/19} \left(\frac{121 e_0^2}{304}+1\right){}^{5779/2299} \eta  y_0^{10}}
   \left(1+\frac{29535 e^2}{31046}+\frac{216378255 e^4}{258556864}+\frac{18164430195 e^6}{24087491072}
\right. \nonumber \\ & \left.   
   +\frac{1920648415789053
   e^8}{2782586968637440}+\frac{61197764904992794077 e^{10}}{95629722868556627968}\right)\,.
\end{align}
\end{subequations}
\end{widetext}
Now, since the functions $y_{\kappa}(e)$, $t_{\kappa}(e)$, $l_{\kappa}(e) = \lambda_{\kappa}(e)$ have been determined analytically in terms of $e$, we are able to simply add them to the already-computed (and validated in \cite{my3PN}) expressions for $y_{\ThreePN,\GR}(e)$, $t_{\ThreePN,\GR}(e)$, $l_{\ThreePN,\GR}(e)$, and $\lambda_{\ThreePN,\GR}(e)$. For details on the efficient generation of the model see \cite{mydata}. 
\section{Bayesian Framework}
\label{sec:bayes_frame}
In this section, we review the techniques we use for our Bayesian statistical analysis of the non-GR terms. We wish to gauge when the non-GR terms are detectable and what constraints we can place on these theories and their parameters. These two questions are subtly different. In gauging when the non-GR terms are detectable, we would like to avoid assuming which model (GR or non-GR) is true \emph{a priori}, and instead, let the data decide. To investigate this question, we can then inject synthetic non-GR signals with varying strengths of the non-GR coupling parameter, and attempt to recover them with both a GR and a non-GR model. A trans-dimensional RJMCMC exploration of the likelihood surface then allows us to compute the Bayes factor in favor of the GR model relative to the non-GR model (or viceversa), given the injected data. Such a study then reveals the value of the coupling parameter at which one would be able to tell that a non-GR effect is present in the data. Alternatively, we can investigate what constraints we can place on the coupling parameters of a non-GR theory, if the signals detected are determined to be consistent with GR. For such a study, we can simulate various GR injections and then carry out a parameter estimation study with a non-GR model. An MCMC exploration of the likelihood surface then allows us to compute the marginalized posterior on the non-GR coupling parameter, and from this extract a projected constraint in terms of a confidence region. 

Let us describe each of these approaches in more detail, starting with the calculation of projected constraints in terms of confidence regions. In producing confidence intervals, we first obtain the full joint posterior, $p(\bm{\theta}|d,M)$, which is the probability of the set of parameters $(\bm{\theta})$ in a model to reproduce the given the data $(d)$ with the model $(M)$. For our non-GR model, the parameters are $\bm{\theta} = \lbrace M_c, \eta, e_{\ast}, \mathcal{A}, \kappa \rbrace$ where $M_c$ is the chirp mass, $\eta$ is the dimensionless mass ratio, $\mathcal{A}$ is an overall amplitude depending on source orientation and distance to source, and $\kappa$ is the non-GR parameter introduced in Sec.~\ref{sec:phasing}, which can be mapped to either the BD or EdGB theory. The eccentricity parameter $e_{\ast}$ sets both $e_0$ and $p_0$ (or equivalently $e_0$ and $y_0$). This parameter is defined to be the eccentricity when the binary has a semilatus rectum corresponding to that of a circular binary emitting GWs at 10Hz. An eccentric signal is composed of many harmonics emitting at several Fourier frequencies at any given time, so a single eccentricity cannot be identified with a single Fourier frequency for all harmonics. As a result, we choose to parameterize our model in terms of $e_{\ast}$, which specifies a reference ellipse from which we can compute the relevant frequency response for each harmonic. For more discussion on the parameter $e_{\ast}$ see \cite{mydata}.

An MCMC algorithm produces samples of the joint posterior, $p(\bm{\theta}|d,M)$, via a random walk through parameter space. Proposed ``steps" for the sampler are drawn from a proposal distribution. The proposed jump is accepted based on a transition kernel, which depends on the prior distribution and the likelihood evaluated both at the current location in parameter space and at the proposed location. Here the likelihood is designed to be consistent with Gaussian noise when evaluated at the correct model that reproduces the signal exactly \cite{1999gr.qc.....3107F}:
\begin{equation}
\mathcal{L}(\theta) \sim \exp \left\lbrace -\frac{1}{2}(d - h(\theta)|d - h(\theta)) \right\rbrace \, ,
\end{equation}
where the inner product between signals is given by
\begin{equation}
(h_1|h_2) = 4\text{Re} \int \frac{\tilde{h}_1^{\ast}\tilde{h}_2}{S_n(f)}df \, .
\end{equation}
In the above ``Re" stands for the real part operator, the overhead tilde stands for the Fourier transform, and $S_n(f)$ is the spectral noise density of the detector. In this work we use spectral noise of aLIGO at designed sensitivity (zero-detuned, high-power), and assume stationary, Gaussian noise when computing the likelihood. We also maximize the likelihood over time and phase constants ($t_c, l_c, \lambda_c$). The signal to noise ratio (SNR) of a signal $h$ is given by $\text{SNR}^2 = (h|h)$. 

In order to produce a one dimensional posterior, which represents the probability that a given parameter $\theta^i$ in a given model can reproduce the data ($p(\theta^i)$), we marginalize the joint posterior over all other parameters, 
\begin{equation}
p(\theta^i) = \int p(\bm{\theta}|d,M) \prod_{k \neq i} d\theta^k \, .
\end{equation}
In order to produce confidence intervals that set an upper limit on a parameter $\theta^i$ we integrate the probability density, $p(\theta^i)$ from $\theta^i = 0$ to an upper bound $\theta^i = \theta^i_{\ast}$ which is determined by requiring a cumulative probability $\Delta$:
\begin{equation}
\label{eq:conf-int}
\int_{0}^{\theta^i_{\ast}} p(\theta^i) = \Delta \, ,
\end{equation}
where $\Delta$ is the desired probability interval (e.g.~for a $90\%$ confidence interval, $\Delta = 0.9$). 

Given all of this machinery, we can now explain how we obtain our confidence intervals to estimate projected constraints on a non-GR parameter. First, we use a signal, $d$, synthesized within GR ($\kappa = 0$), and use an MCMC to explore the parameters of a non-GR model ($\kappa \neq 0$). This exploration leads to a joint posterior distribution, from which we can compute the marginalized posterior on $\kappa$, $p(\kappa)$. This marginalized posterior will be centered at zero, and we can then use it to produce upper bounds via confidence intervals through Eq.~\eqref{eq:conf-int}. Confidence-interval constraints on $\kappa$ can then be mapped to projected constraints on the BD and EdGB coupling parameters. This is a standard technique to produce upper limits on modified gravity  theories~\cite{2019arXiv190304467T, 2019arXiv190807089M} (though this is often done with a Fisher analysis in place of a full MCMC implementation and analysis).

In producing our confidence intervals we use a MCMC which we have written from scratch in \texttt{C++}. Our proposal distribution consists of a mix of draws from the prior distributions, Fisher jumps, and differential evolution jumps. We also employ parallel tempering to ensure efficient and complete exploration of the posteriors by the MCMC sampler. We start the MCMC at the maximum likelihood solution, so in principle there is no burn in phase, but we still choose to discard the first 1000 samples. We have verified that the MCMC has converged by taking different subsets of samples and verifying the the end results do not change. Since the MCMCs are run for a set period of time on our computing cluster as opposed to a set number of iterations, the total number of iterations can vary from run to run. Generally they generate a total of $\sim 200,000$ samples of the posterior, which we have ensured is more than enough for the chains to have converged.

In the above analysis, we assume that the signal is described by GR, but use a non-GR model to extract it, so as to determine how large the non-GR effects can be, while remaining consistent with statistical noise. We now also want to investigate a separate question: how large of a non-GR deviation is needed in the signal so that the data itself supports a non-GR model over a GR model. To investigate this question, so we must consider the ratio of the probability of the two models given the data. From Bayes' Theorem, the probability of model $i$ given the data is related to the likelihood via 
\begin{equation}
p(M_i|d) = \frac{p(M_i|I)p(d|M_i)}{p(d)} \,
\end{equation}
where $p(d|M_i)$ is the likelihood for model $i$, $p(d)$ is the evidence, and $p(M_i|I)$ is the prior. The ratio of probabilities of two models is the odds ratio, namely
\begin{equation}
\label{eq:bayes_fact}
\mathcal{O}_{1,2} = \frac{p(M_1|I)}{p(M_2|I)}\frac{p(d|M_1)}{p(d|M_2)} = \frac{p(M_1|I)}{p(M_2|I)}\mathcal{B}_{1,2} \, ,
\end{equation}
where in the ratio the model evidences have canceled out. The quantity $\mathcal{B}_{1,2}$ is the Bayes factor, which in this work will be equal to the odds ratio as we assume that the prior probability for either model is the same. The Bayes factor represents the level of evidence for model 1 versus model 2. Different values for the Bayes factor represent different levels of belief in either model 1 or 2, as shown in Table \ref{tab:Bayes}.
\begin{table}
\begin{tabular}{||c|c|c||}
\hline
$\log_{10}{\mathcal{B}_{1,2}}$ & $\mathcal{B}_{1,2}$ & Evidence \\
\hline
0 - 1/2 & 1 - 3.2 & Inconclusive \\
\hline
1/2 - 1 & 3.2 - 10 & Substantial \\
\hline
1 - 2 & 10 - 100 & Strong \\
\hline
 $>$ 2 & $>$ 100 & Decisive \\
\hline
\end{tabular}
\caption{\label{tab:Bayes} Bayes factors and their associated level of evidence in favor of Model 1.}
\end{table}

To better understand the Bayes factors, we invoke a standard derivation, presented for example in \cite{Gregory:2005:BLD:1051497}, wherein we write the global likelihood as 
\begin{align}
\label{eq:model_like}
p(d,M_i) &= \int p(\bm{\theta}|M_i,I)p(d|\bm{\theta},M_i) d\bm{\theta} \, \nonumber \\
& = \int p(\bm{\theta}|M_i,I)\mathcal{L}(\bm{\theta})  d\bm{\theta} 
\approx \mathcal{L}(\bm{\hat{\theta}}) \frac{V_{\sigma,\theta}}{V_{\theta}}
\end{align}
where we see in the first two equalities that the global model likelihood is just the likelihood evaluated at the best-fit parameters, weighted by the priors. In the last equality, we have approximated the priors are uniform, and thus, given rise to a factor of the prior volume $V_{\theta}$. We have also approximated the likelihood for the model parameters to be a multivariate Gaussian, leading to the factor of the likelihood evaluated at its maximum, $\mathcal{L}(\bm{\hat{\theta}})$, and $V_{\sigma,\theta}$ which is the 1-sigma posterior volume (since the priors are assumed to be flat, the 1-sigma posterior volume is equivalent to the 1-sigma likelihood volume). It is then straightforward to approximate the Bayes factor from Eq.~\eqref{eq:bayes_fact} using the above, where we assume that we have a nested model with model 1 including an extra parameter $\kappa$ to specialize to the case we consider in this work, 
\begin{equation}
\label{eq:bayes_app}
\mathcal{B}_{1,2} \approx \frac{\mathcal{L}(\bm{\hat{\theta}}_1)}{\mathcal{L}(\bm{\hat{\theta}}_2)} \frac{\delta \kappa}{\Delta \kappa} \, ,
\end{equation}
where $\delta \kappa$ is a characteristic width of the marginalized posterior of $\kappa$ and $\Delta \kappa$ is the prior volume of the prior on $\kappa$. We note that since model 1 contains model 2 as a special case, the first factor can never be less than unity, i.e.~the ratio of maximum parameter likelihoods will always favor the more complicated model. However the second factor, called the ``Occam factor,'' penalizes the more complicated model for ``wasted" parameter space. The consequence of Eqs.~\eqref{eq:bayes_app} and~\eqref{eq:model_like} is that the Bayes factor is sensitive to the prior volume and shape.

While the above form is a useful approximation to understand the main features that affect the Bayes factor, we choose to compute the Bayes factor via a trans-dimensional reversible jump MCMC \cite{Green95reversiblejump}. Within a given model, we use Fisher jumps, draws from the prior distribution, and differential evolution. For trans-dimensional jumps between models, we use draws from a log flat distribution on $\kappa$ (or whichever modified theory parameter is being explicitly jumped in). Parallel tempering is used to ensure a complete and efficient exploration of the posteriors; we find that most of the trans-dimensional jumps in the $T = 1$ chain are from parallel tempering. Our prior distributions are as follows: $\log{\mathcal{M}_c} \sim \text{U}[-0.22, 2.3]$, $\eta \sim \text{U}[0.08, 0.25]$, $e_{\ast} \sim \text{U}[0, 0.805]$, and $\log{\mathcal{A}} \sim \text{U}[-46, -37]$. As we have made clear that the prior on the alternative theory parameter can affect the Bayes factor heavily, we explore two different priors in the next section: a flat prior on $\kappa$ and a flat prior on $\alpha^2$ as this is what enters the waveform linearly (a flat prior on $\kappa$ implies a non-trivial prior on $\alpha^2$ and vice versa).
\section{Results}
\label{sec:results}
\begin{figure*}[htp]
\includegraphics[clip=true,angle=0,width=0.475\textwidth]{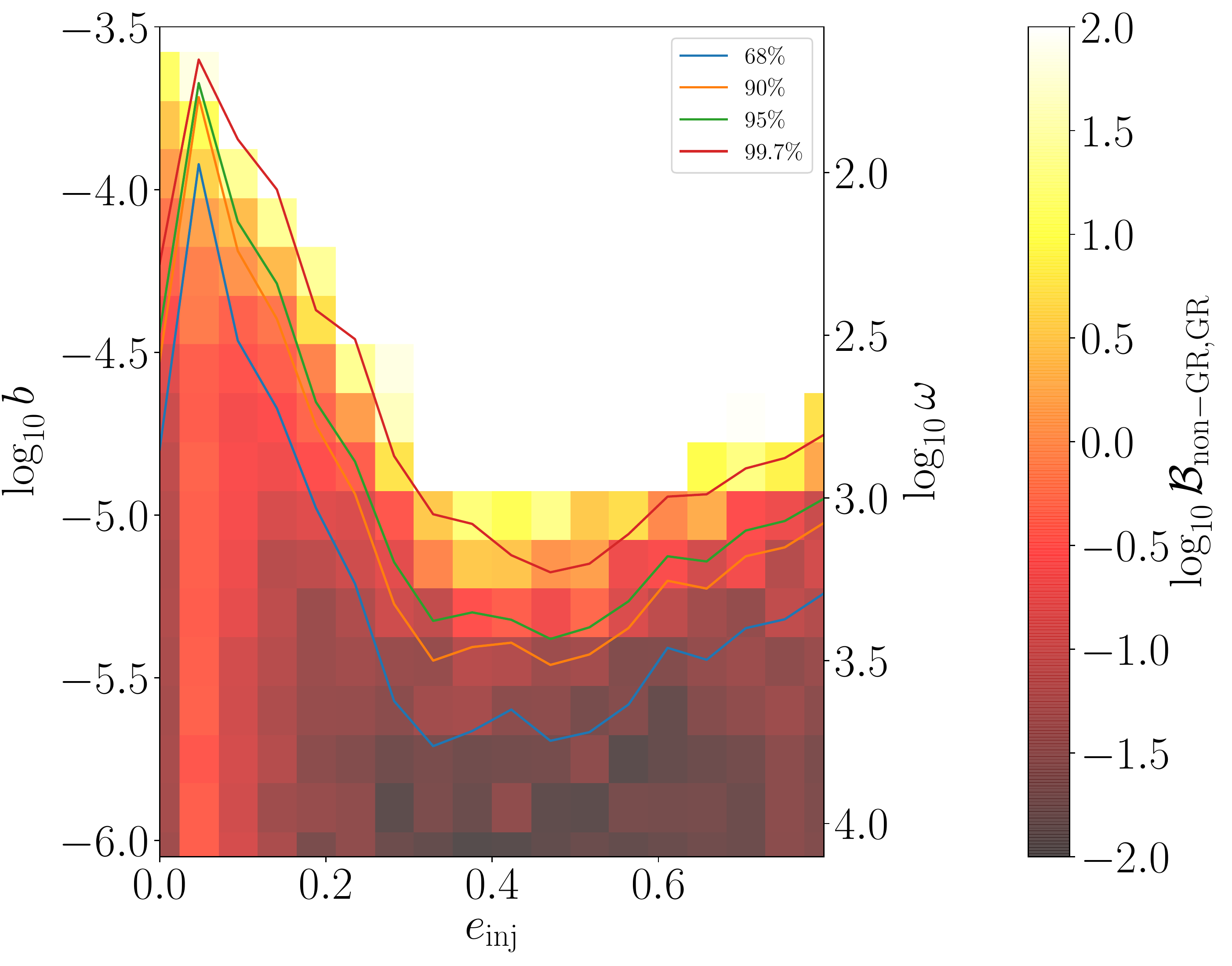}
\qquad 
\includegraphics[clip=true,angle=0,width=0.475\textwidth]{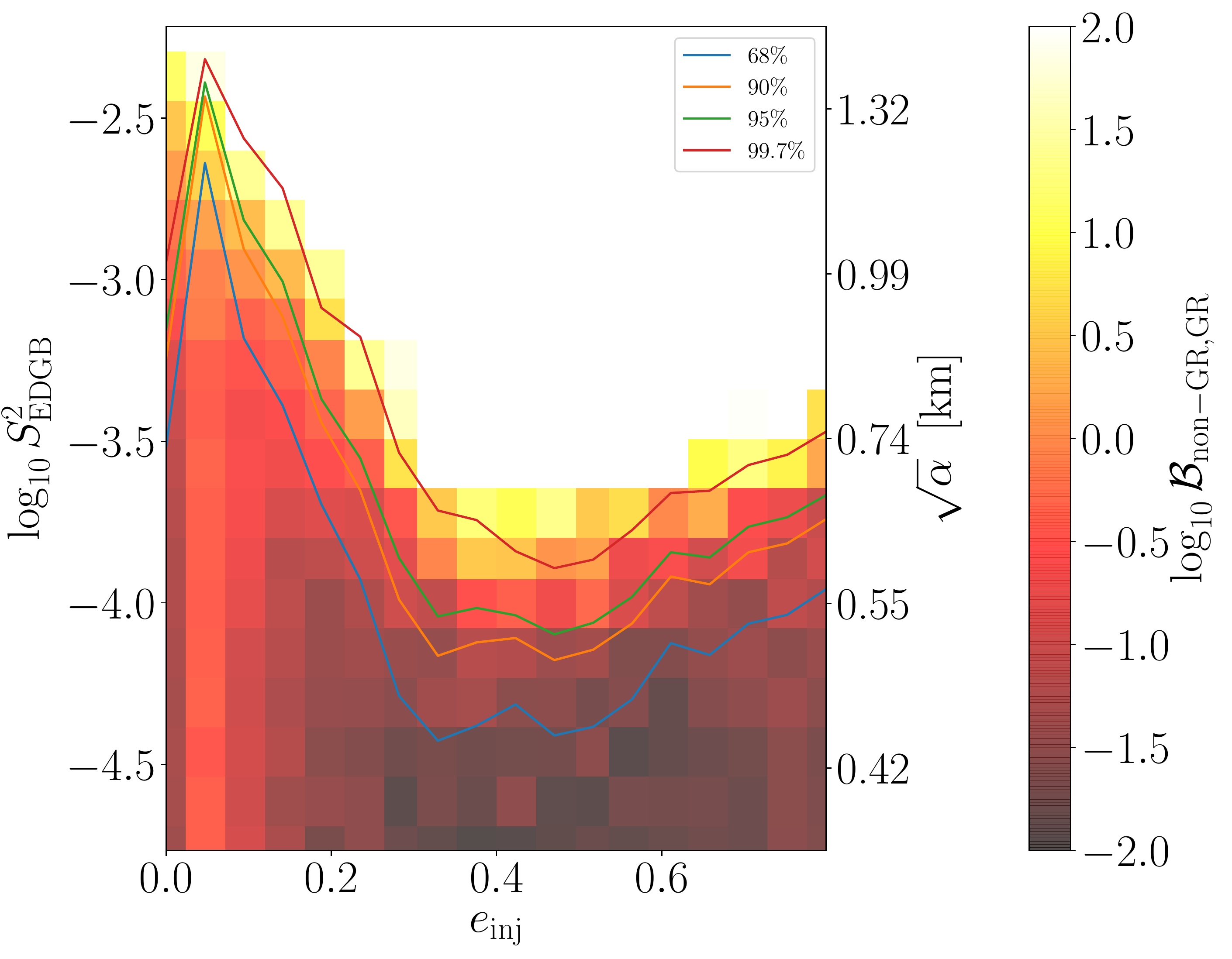}
\caption{\label{fig:bh10_flat_alpha} The left (right) panel shows the Bayes factor in favor of the Brans-Dicke (EdGB) model over the GR model as a function of injected eccentricity and coupling parameter, using a flat prior on $\kappa$. The lines indicate upper limits on the coupling parameter for the 68$\%$, 90$\%$, 95$\%$, and 99.7$\%$ confidence regions. In the Bayes factor analysis, we see an enhancement of about a factor of $10^{3/2}$ in the sensitivity to the coupling parameter. In the confidence interval analysis, we see an enhancement of of a little under 10 in the ability to constraint the coupling parameter.}
\end{figure*}
In this section we present the results of our upper bound analysis on the modified theory coupling parameter, along with our Bayes factor analysis. In our Bayes factor analysis, we grid the $(\kappa, e_{\ast})$ parameter space and inject signals on this grid assuming a $(m_1, m_2) = (1.4, 10)M_{\odot}$ inspiral with an SNR of 30. In our upper bound analysis, we inject data that is consistent with a GR waveform ($\kappa = 0$) and conduct parameter estimation with a non-GR model to estimate the marginalized posterior on $\kappa$ for eccentric inspirals with the same eccentricities used in the Bayes factor analysis. We then translate our marginalized posteriors on $\kappa$ to posteriors on the BD parameters ($b$, $\omega_{\BD}$) and the EdGB parameters ($\mathcal{S}^2_{\EdGB}$, $\alpha$). In relating $b$ to $\omega_{\BD}$ we have assumed that the $1.4 M_{\odot}$ object is a neutron star with a sensitivity $s = 0.171$, which is what one would find for an APR equation of state \cite{1998PhRvC..58.1804A, 1975ApJ...196L..59E}. We also assume the $10 M_{\odot}$ object is a black hole with $s = 0.5$ by the no hair theorems~\cite{Hawking:1972qk}. In the EdGB case, we again assume the binary is composed of a neutron star and a black hole, and thus, we have that $\mathcal{S}^2_{\EdGB} = \xi/m^4_{\BH} = 16 \pi \alpha_{\EdGB}^{2}/m^{4}$.

\subsection{What constraints can we put on non-GR theories with eccentric signals?}

In Figure \ref{fig:bh10_flat_alpha} we present both the Bayes factor as a function of the values of the injected modified theory parameter indicated on the y-axes and the injected eccentricity. We also overplot contours that represent the upper bound on the modified theory parameter given certain confidence intervals as a function of the injected eccentricity. As the injected eccentricity is raised from $e_{\rm inj} = 0$ to $e_{\rm inj} \sim 0.4$, the region in coupling parameter where the Bayes factor begins to provide significant support for the modified theory model is lowered by a factor of about $10^{3/2}$. To be clear, by $e_{\rm inj}$ we here mean the injected value of $e_{\ast}$ in the signal. In the confidence interval curves, we see that for small eccentricities ($e_{\rm inj} \leq 0.2$) the upper bound on the coupling parameter is less stringent or about as stringent as in the circular case. Still however, as the eccentricity is raised to $e_{\rm inj} \sim 0.4$, the constraint improves by about an order of magnitude. 

We can now compare these projected constraints to current constraints. Recall from Sec.~\ref{intro} that the current constraint on EdGB gravity from the GW170608 observation is $\sqrt{\alpha} < 5.6$km at $90\%$ confidence, while the current constraint on BD theory from the Cassini mission is $\omega > 40,000$ at $95\%$ confidence. Taking the minimum of the appropriate confidence interval from Fig.~\ref{fig:bh10_flat_alpha} suggests that one could obtain GW constraints as good as $\sqrt{\alpha} \lesssim 0.5$km at $90\%$ confidence and $\omega \gtrsim 2700$ at $95\%$ confidence. This is an improvement of roughly an order of magnitude in the EdGB case, but it does not improve the current constraint on the BD coupling parameter. In this analysis, we have assumed a uniform prior on the coupling parameter $\kappa \sim \text{U}[0, 0.034]$; our constraints would be even more stringent if we have chosen a uniform prior on the logarithm of the coupling parameter, as this would put more wait around GR values.  

Although this projected constraints are interesting, it is important to remember that the analysis here assumes an SNR of 30 in the inspiral only (whereas GW170608 had an SNR of 13 for the entire waveform). Such an increase in SNR will be possible once LIGO reaches designed sensitivity for a sufficiently nearby source. More dangerous perhaps is the use of a significantly reduced set of parameters, importantly neglecting spin effects. The modifications we consider here, however, are -1PN order, while spin effects start at +1.5PN order in the waveform phase. For this reasons, covariances between the modified gravity effects we consider here and spin effects should be very weak and not affect our conclusions significantly.

There does, however, arise a complication in the mapping between $\mathcal{S}^2_{\rm EdGB}$ and $\sqrt{\alpha}$ (and thus between $\kappa$ and $\sqrt{\alpha}$) when spins are considered. As discussed extensively in \cite{Yunes:2016jcc, quadratic} (see e.g.~Appendix D in~\cite{Yunes:2016jcc}), the spin dependence of $\mathcal{S}^2_{\rm EdGB}$ makes it impossible to constrain $\sqrt{\alpha}$ unless the spins are well measured. This is because there are values of the spins and the masses for which $\mathcal{S}^2_{\rm EdGB}$ vanishes identically irrespective of the value of $\sqrt{\alpha}$. Therefore, if the masses and spins cannot be constrained well enough to disallow this possibility, then $\sqrt{\alpha}$ cannot be constrained. This effect can be mitigated by future detectors which will be able to resolve the spin of the compact objects better than current detectors. 

Our projected BD constraint is stronger than what one finds in the quasi-circular limit, but still not stronger than current constraint from the Cassini spacecraft. This suggests that in order to obtain a better constraint, even with the most optimal system configuration, we will likely need to wait for more sensitive ground based detectors. For example, in Fig.~5 of \cite{2019arXiv190807089M}, a Fisher matrix analysis is used to project constrain on $\omega$ using GWs from from binaries with small eccentricity. Their work suggests that Cosmic Explorer and Einstein telescope could supersede the current constraint on BD, especially if their results are extrapolated to our moderate eccentricity results. 

Why does the constraint on the non-GR parameter deteriorate for eccentricities past $0.4$? Recall from Eq.~\eqref{eq:bayes_app} that the Bayes factor scales as the ratio of the maximum likelihoods for either model times the ratio of the characteristic width of the posterior on the non-GR parameter to the prior width. As the eccentricity is increased from 0.4 to 0.8, the ratio of the maximum likelihoods decreases in magnitude at the same rate as the Bayes factor decreases. By Eq.~\eqref{eq:bayes_app}, this suggests that as the eccentricity is increased from 0.4 to 0.8, the width of the posterior on the non-GR parameter remains roughly constant, which implies the likelihood is less sensitive to the modified theory parameter; note that this effect can also be seen in the other system parameters, as shown e.g.~in \cite{mydata}. This would also explain the deteriorated constraint from the confidence intervals. 

But what is happening to the waveform and the likelihood to cause this effect? As the inner product that enters likelihood calculation, $(d - h(\theta)|d - h(\theta))$ , is highly sensitive to phase differences between the data $(d)$ and the template $(h)$, the overall number of phase cycles is a useful metric to understand how sensitive the likelihood can be to the system parameters. If there are many phase cycles, then we might expect that a small change in parameters could (over these many cycles) cause a significant dephasing. However, if the number of phase cycles is small then a change in the parameters of the model may not lead to enough dephasing to significantly affect the inner product entering the likelihood.

Since eccentricity enhances energy and angular momentum loss, it increased eccentricity forces the binary to inspiral faster, and thus, to produce less cycles of phase. However, the higher the eccentricity, the more important higher harmonics are, and the latter can produce even more cycles of phase, as they emit at higher multiples of the orbital frequency. Thus we expect that there is some interplay between these two contributions to the number of phase cycles, and ultimately, this quantity should be computed in order to understand the loss in likelihood sensitivity to parameter changes.

To do so, we first extend the circular calculation of the number of cycles of phase to the case of many harmonic applicable to eccentric inspirals. In the circular case, the number of phase cycles is given by~\cite{2005CQGra..22S.801D}
\begin{equation}
\label{eq:circ_num}
\Delta N_{\psi} = \frac{1}{2\pi}\left[ \left. \psi(f_2) - \psi(f_1) + (f_1 - f_2)\frac{d\psi}{df}\right|_{f = f_1}\right]  \, ,
\end{equation}
where $f_1$ is the initial frequency, $f_2$ is the final eccentricity, and the derivative term ensures that $\Delta N_{\psi}$ and its frequency derivative vanish at $f_1$. In the eccentric case, the presence of many harmonics complicate calculations of the \it overall \rm phase, so it is preferable to work with the individual phases of each harmonic, $\psi_j$. Each harmonic $j$ contributes some number of phase cycles:
\begin{equation}
\Delta N_{\psi_{j}} = \frac{1}{2\pi}\left[ \left. \psi_j(f_2) - \psi_j(f_1) + (f_1 - f_2)\frac{d\psi_j}{df}\right|_{f = f_1}\right] \, .
\end{equation}
In order to find the total number of phase cycles we choose to weight each harmonic contribution by its fractional SNR. Thus, for the total number of phase cycles we now have 
\begin{equation}
\label{eq:ecc_num}
\Delta N_{\psi} = \frac{1}{(h|h)^{1/2}}\sum_j (h_j|h_j)^{1/2} \Delta N_{\psi_{j}} \, .
\end{equation}
Here $h$ is the full signal given by Eq. \eqref{eq:spa_simp}, and the individual harmonics, $h_j$, are given by individual terms in the sum on $j$ in Eq. \eqref{eq:spa_simp}. Note that this reduces to Eq. \eqref{eq:circ_num} in the circular limit.
\begin{figure}[htp]
\includegraphics[clip=true,angle=0,width=0.475\textwidth]{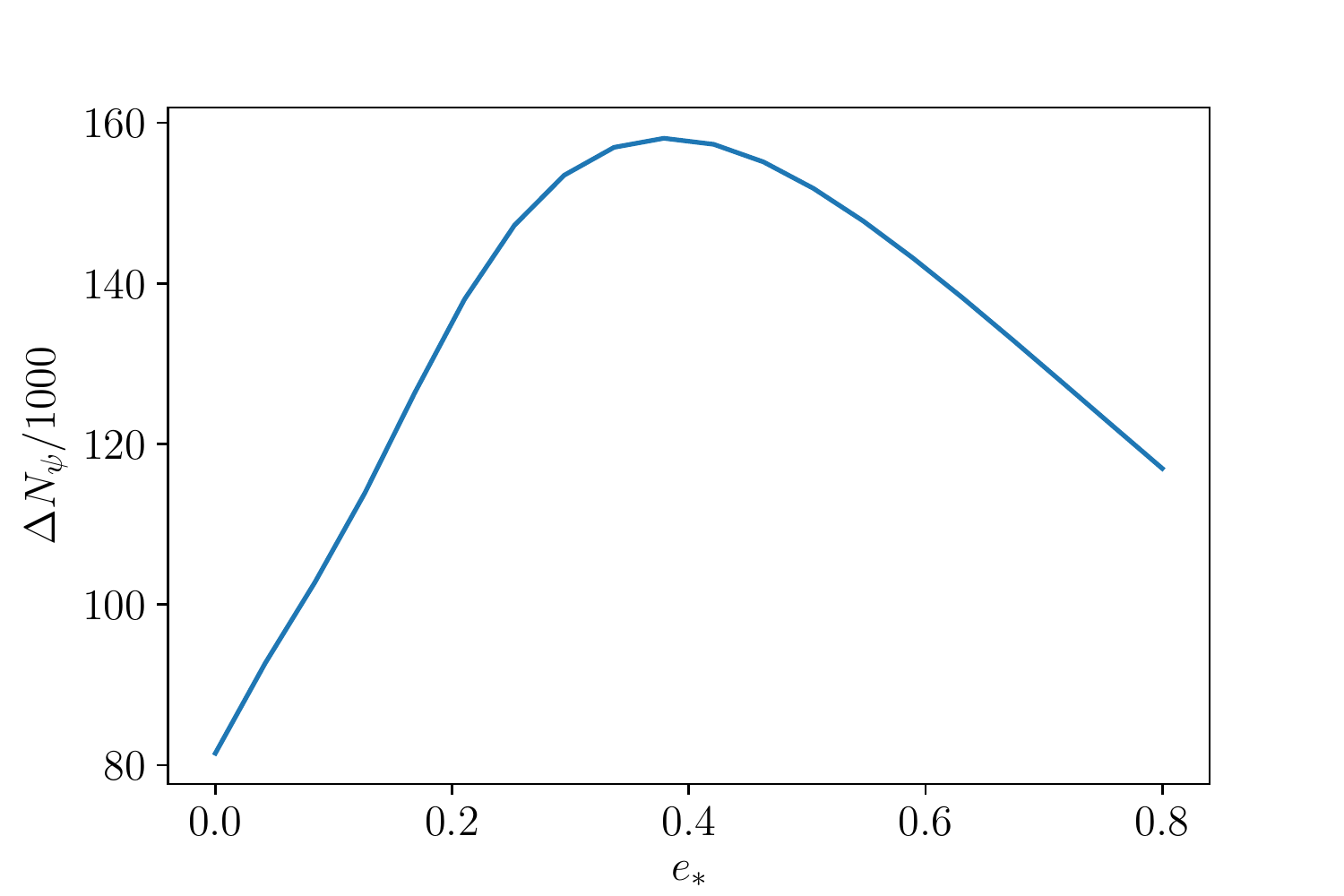}
\caption{\label{fig:phase_cyc} The number of phase cycles as given by Eq. \eqref{eq:ecc_num} as a function of the eccentricity. Initially, the number of phase cycles increases until eccentricities $\sim 0.4$ at which point they begin to decrease.}
\end{figure}

\begin{figure*}[htp]
\includegraphics[clip=true,angle=0,width=0.475\textwidth]{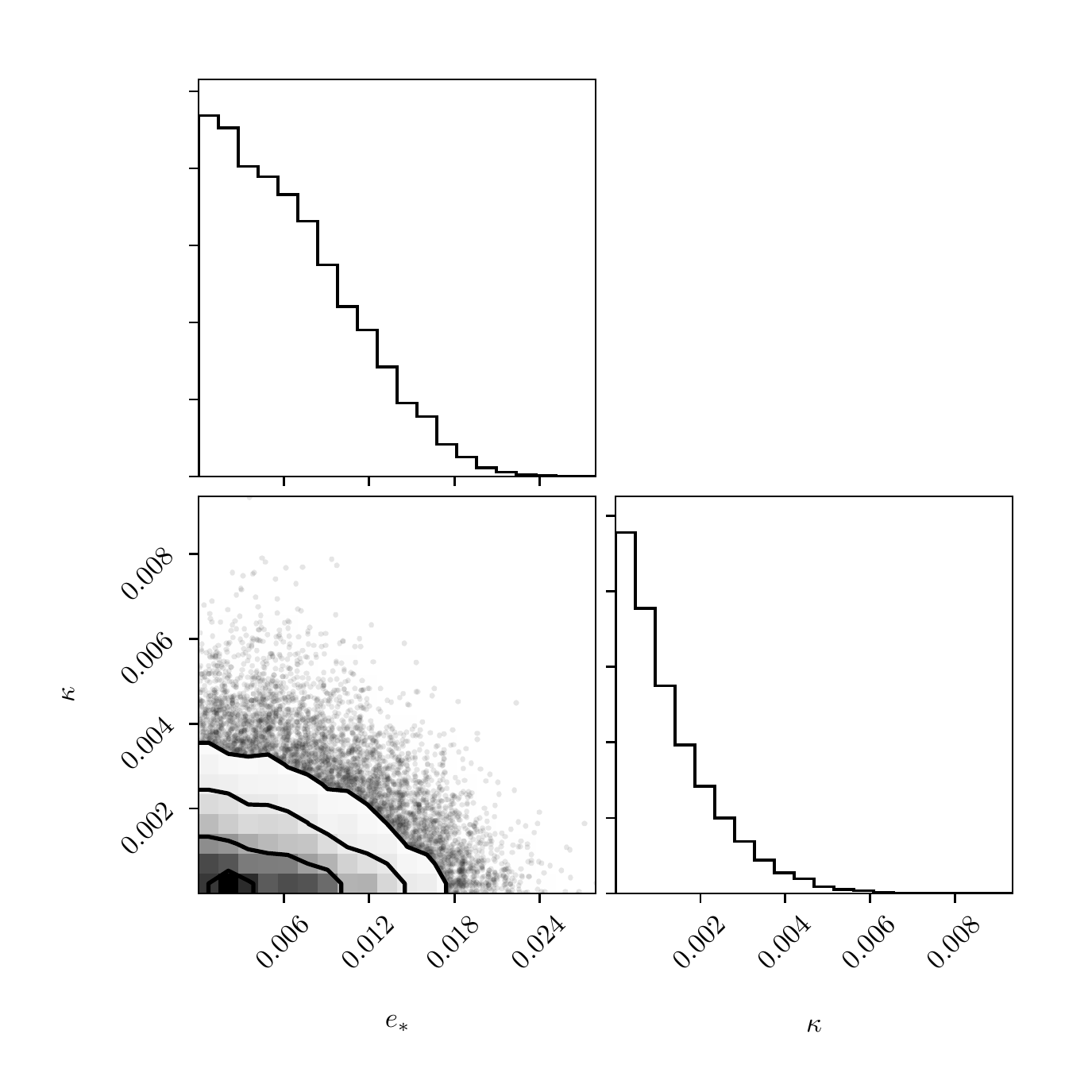}
\includegraphics[clip=true,angle=0,width=0.475\textwidth]{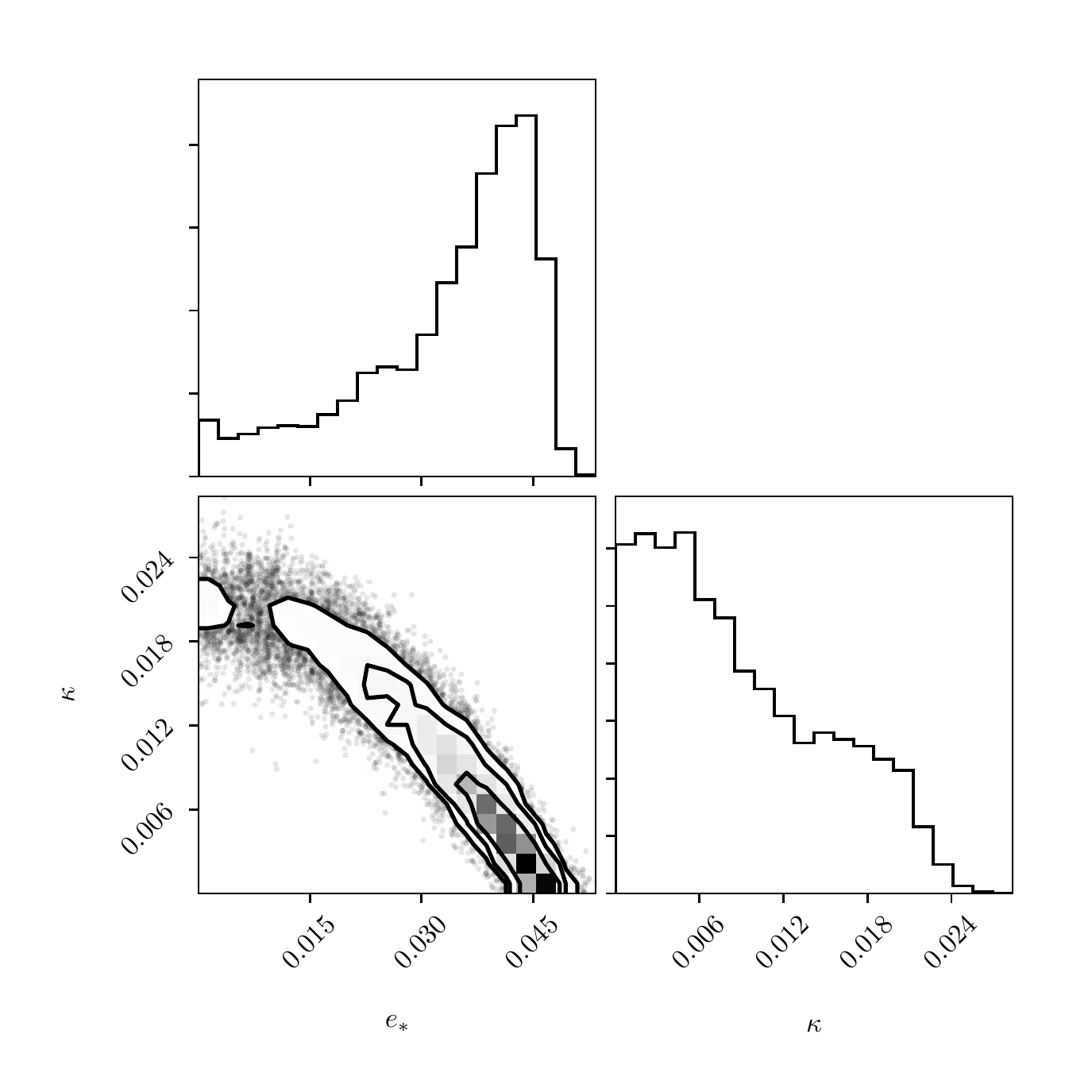}
\caption{\label{fig:covar} Corner plots for the posteriors on $e_{\ast}$ and $\kappa$ when injecting a GR signal and recovering with a modified theory model. The injected eccentricity is $0$ (left) and $0.047$ (right). For small eccentricities, a large negative covariance arises.}
\end{figure*}
In Figure \ref{fig:phase_cyc} we plot the weighted phase cycles (Eq~\eqref{eq:ecc_num}) as a function of the eccentricity. Note that as suspected, we see that there is an interplay between the binary inspiraling faster and producing less phase cycles, and the increasing number of harmonics contributing more phase cycles. Up until eccentricities near $0.4$, the number of cycles increases, but then for higher eccentricities the number of phase cycles decreases. The (inverted) shape is nearly identical to the shape of the confidence interval constraints on Fig.~\ref{fig:bh10_flat_alpha}. This is strong evidence that the source of the likelihood's decreased sensitivity to parameters is the decreasing number of phase cycles. This in turn sources a deterioration of the constraints on the non-GR parameters, as well as on the source parameters as pointed out in~\cite{mydata}, as the eccentricity is increased above $\sim 0.4$.

\subsection{Disagreement at circularity}
Interestingly the upper bound produced by the confidence interval very closely follows contours of the Bayes factor, except in the circular case. In order to understand this behavior, it is helpful to look at the covariance between $e_{\ast}$ and $\kappa$, as well as the marginalized posterior in $\kappa$. Figure \ref{fig:covar} shows the covariance between these parameters, as well as their marginalized posterior for the $e_{\rm inj} = 0$ and $e_{\rm inj} = 0.047$ cases. We see that for small eccentricities there arises a strongly negative covariance between these parameters. In turn, this covariance leads to the exploration of much larger values of the modified theory coupling parameter, giving a less stringent bound on the parameter when $e_{\rm inj} = 0.047$. This covariance is sensible as the eccentricity and the non-GR parameter \emph{both} lead to a similar physical effect: an increased rate of energy and angular momentum loss by the binary, and therefore a faster inspiral. As the injected eccentricity becomes large, however, this covariance disappears, likely because the effect of eccentricity begins to introduce considerable more power in higher harmonics, which cannot be duplicated by the modified theory parameter. 

In the case of the Bayes factors, we do not see an increased constraint at circularity. The key to this feature is that, unlike in the confidence interval analysis, the signal has a non-zero value of the non-GR parameter. As such, even when $e_{\rm inj} = 0$, the GR model is free to explore higher eccentricities along the covariance between $\kappa$ and $e_{\ast}$. In exploring that covariance, the GR model can achieve a higher maximum likelihood and thus inflate the denominator in Eq.~\eqref{eq:bayes_app} leading to smaller Bayes factors. To summarize, in the confidence interval analysis, the covariance between $e_{\ast}$ and $\kappa$ is small when both injected values are zero, however in the Bayes factor analysis even when $e_{\rm inj} = 0$ the injected value of the non-GR parameter is nonzero, so the effects of this covariance ``turning on" are not seen. 

\subsection{How sensitive are the results to the priors?}
\begin{figure*}[htp]
\includegraphics[clip=true,angle=0,width=0.475\textwidth]{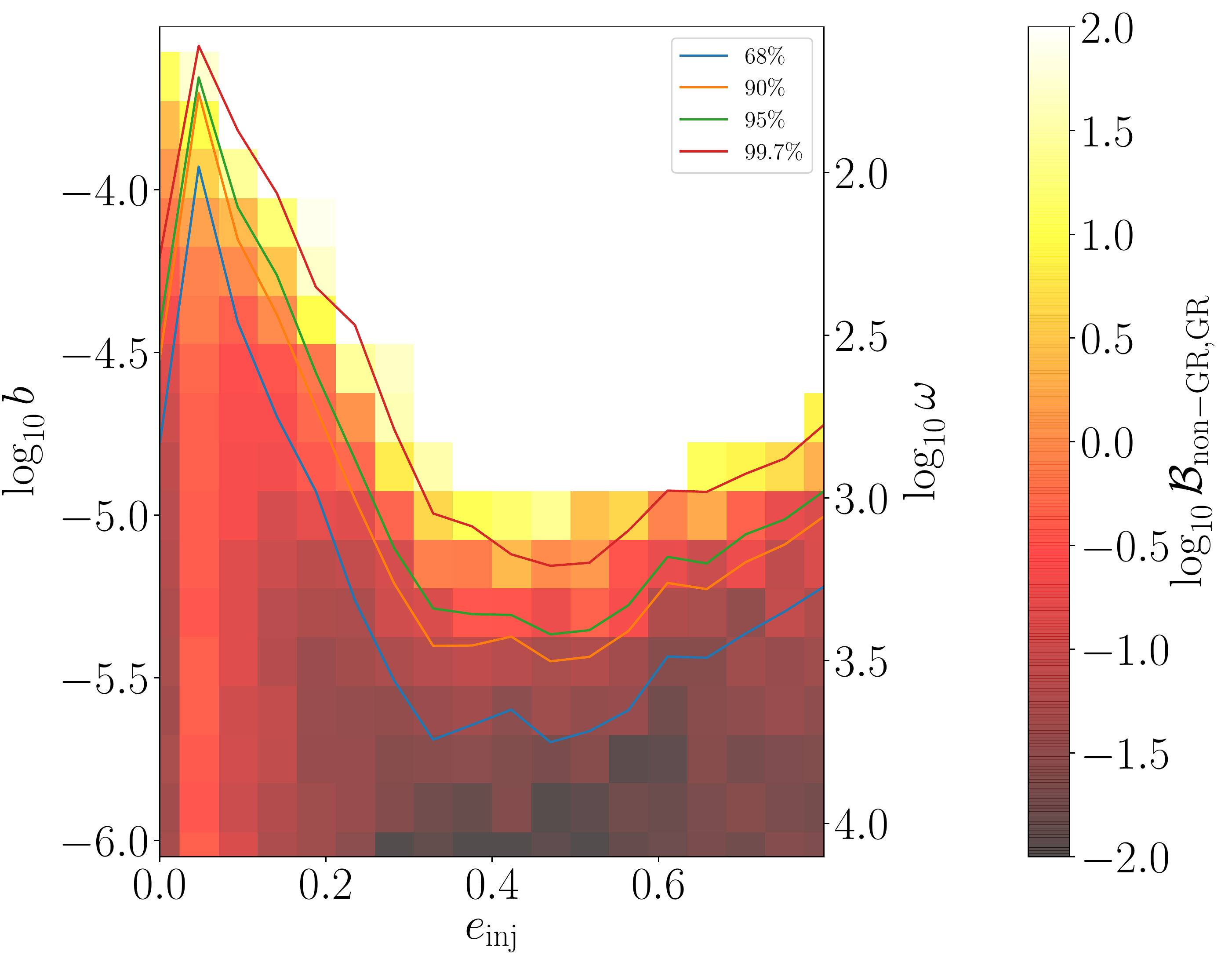} 
\qquad 
\includegraphics[clip=true,angle=0,width=0.475\textwidth]{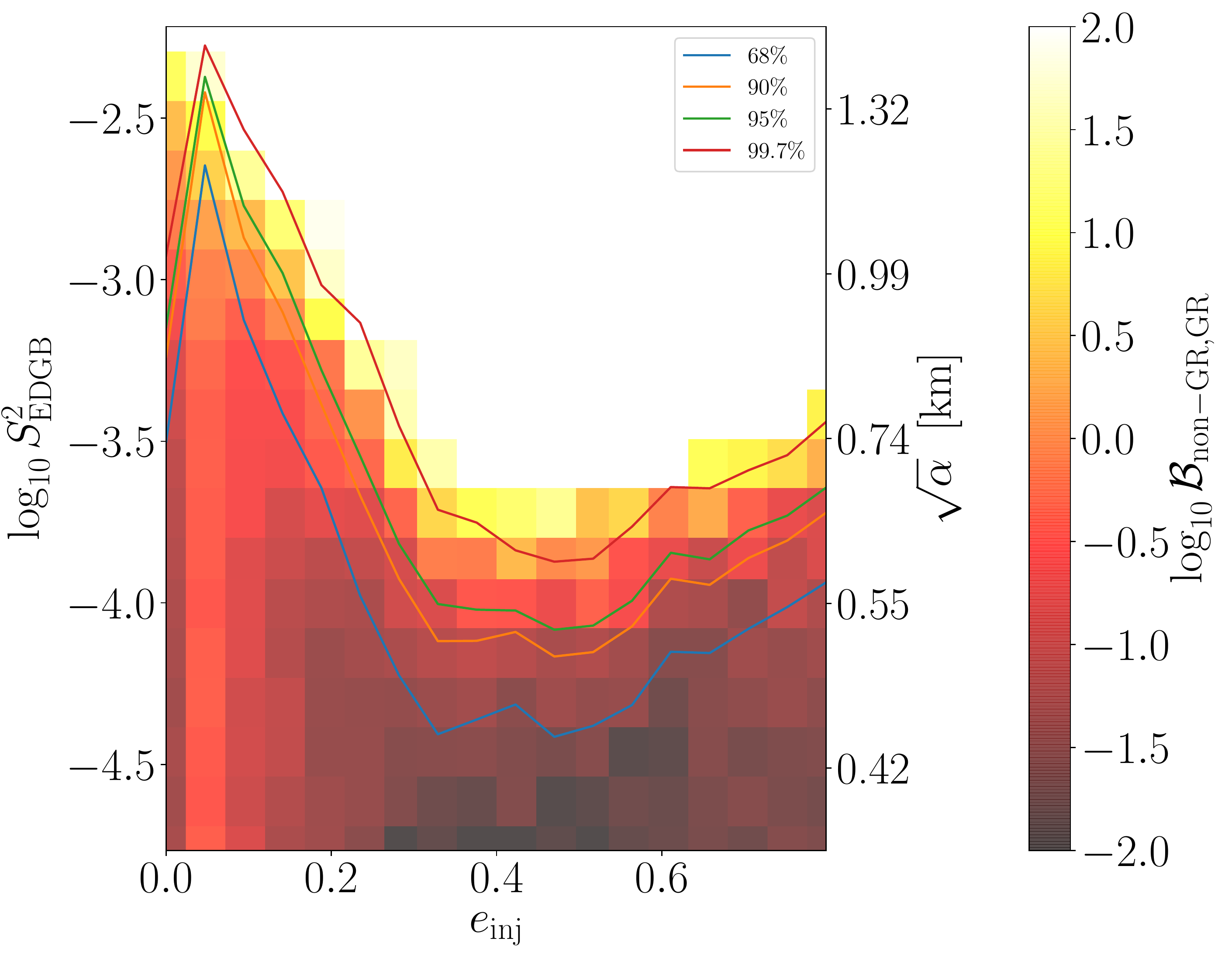}
\caption{\label{fig:bh10_flat_xi} Same as Figure \ref{fig:bh10_flat_alpha}, but here we have used a flat prior in $\alpha^2$}
\end{figure*}
\begin{figure}[htp]
\includegraphics[clip=true,angle=0,width=0.475\textwidth]{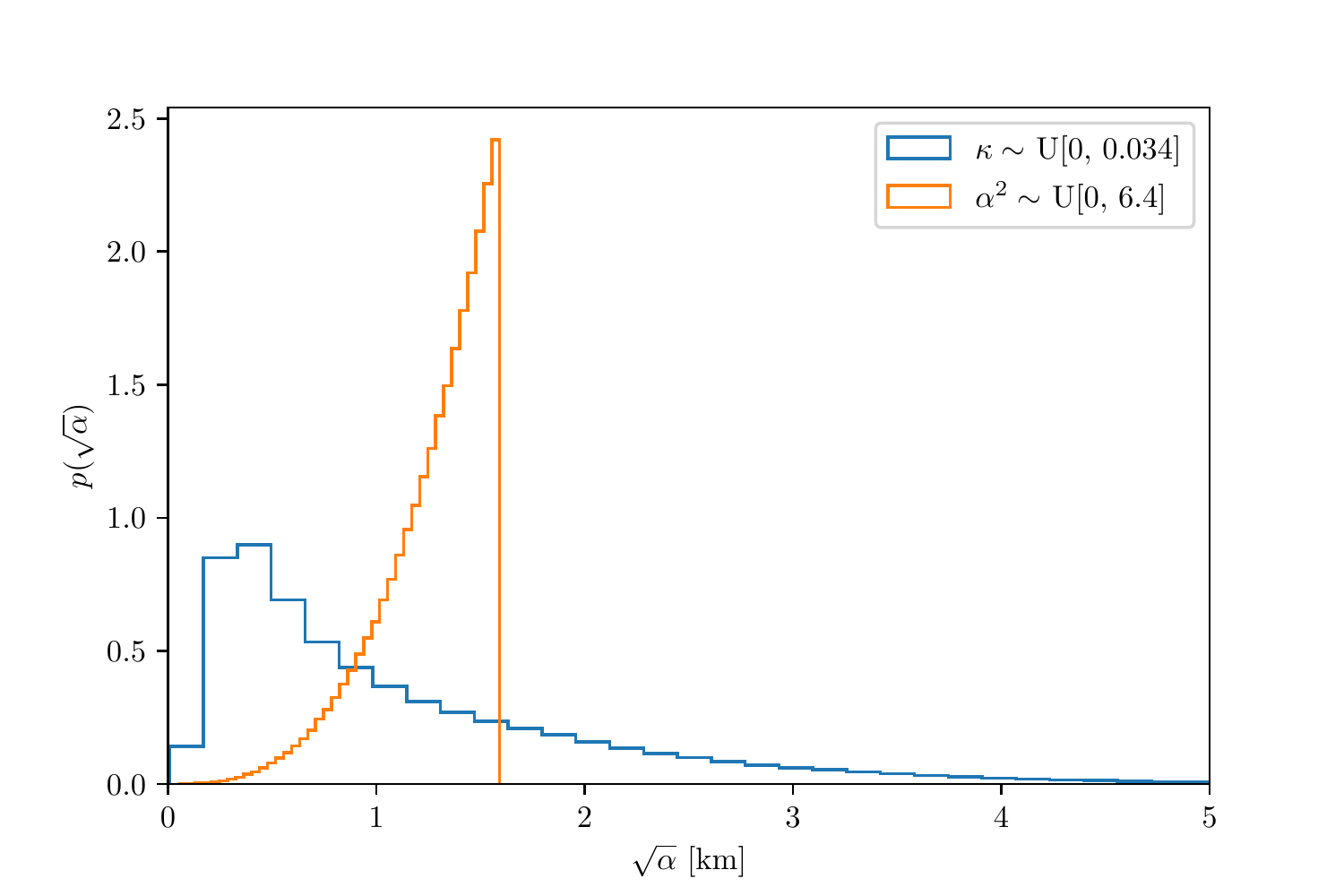}
\caption{\label{fig:priors} The prior probability distributions for $\sqrt{\alpha}$ when the prior on $\kappa$ is flat (blue) and when the prior on $\alpha^2$, which enters the phasing linearly, is flat (orange).}
\end{figure}

Since the Bayes factor is highly sensitive to the prior volume as well as the shape of the prior (see e.g. Eqs.~\eqref{eq:model_like} and~\eqref{eq:bayes_app}), it is worth exploring what the prior looks like when converted from $\kappa$ to the modified theory coupling constant. In Fig.~\ref{fig:bh10_flat_alpha} we chose a uniform prior on $\kappa$, i.e.~$\kappa \sim \text{U}[0, 0.034]$, so let us explore how this translates to a prior on the EdGB parameter $\sqrt{\alpha}$, which we show in Fig.~\ref{fig:priors}. Clearly, a uniform prior on $\kappa$ leads to a non trivial prior on $\sqrt{\alpha}$. We also show a flat prior on $\alpha^2$ whose upper bound is consistent with the upper bound obtained with a uniform prior on $\kappa$ (assuming the same injected mass). We choose here to compare against a uniform prior on $\alpha^2$, instead of a uniform prior on $\sqrt{\alpha}$, because $\alpha^2$ enters the phase functions linearly (while $\sqrt{\alpha}$ is ultimately what is to be constrained). We are then forced to wonder if the results of our Bayes factor analysis would be greatly affected if we explicitly used uniform priors on the coupling parameter $\alpha^2$. This is particularly important because often the widely used ppE parameters do not map linearly to coupling parameters in different theories. A flat prior in those generic corrections in the ppE formalism could imply non-trivial prior on the coupling parameter that is ultimately being constrained. See for example the posteriors on $\sqrt{\alpha}$ obtained through analysis of GW170808 in \cite{2019PhRvL.123s1101N} where a prior much like the one shown in Fig. \ref{fig:priors} enforces nearly zero support in the posterior at $\sqrt{\alpha} = 0$. We have verified that we obtain a similar result in the posterior of $\sqrt{\alpha}$ when using a flat prior on $\kappa$.
  
In Figure \ref{fig:bh10_flat_xi} we show the Bayes factors and confidence intervals, as in Fig.~\ref{fig:bh10_flat_alpha}, but now assuming a uniform prior on $\alpha^2$: $\alpha^2 \sim \text{U}[0, 6.4]$. We see that despite the differently shaped priors, the Bayes factors barely differ quantitatively, and virtually suffer no qualitative differences. We have verified numerically that they are, in fact, slightly different. This is reassuring, since it implies that provided the bounds of the priors are consistent, we see a consistent Bayes factor in this case; the Bayes factors change considerably if we increase our prior volume, but this is a well-known feature of any Bayes factor analysis. We can conclude that it is sufficient to have one parameter estimation or model selection run with a single $\kappa$ coefficient, which can then meaningfully be translated to various coupling parameters, regardless of the non-linearity of the mapping between them implying non trivial priors. This, however, will not always be the case, and the robustness of the calculation will depend on the functional form of the mapping between the ppE parameter and the coupling constant of the theory. For example, Ref.~\cite{Yunes:2016jcc} showed in detail (see Appendix D in that paper) that a well-constrained value of the ppE parameter leads to a wide number of allowed values of $\sqrt{\alpha}$ due to unconstrained spins in the mapping between $\sqrt{\alpha}$ and the ppE parameter\footnote{Particularly, see their Figure 15 where GW150914 is shown to place limits between $\sqrt{\alpha} \sim 0$ and $\sqrt{\alpha} < 40$km depending on the values of the (unconstrained) spins.}. 
\subsection{Uncertainty in Bayes Factor}
\begin{figure}[htp]
\includegraphics[clip=true,angle=0,width=0.475\textwidth]{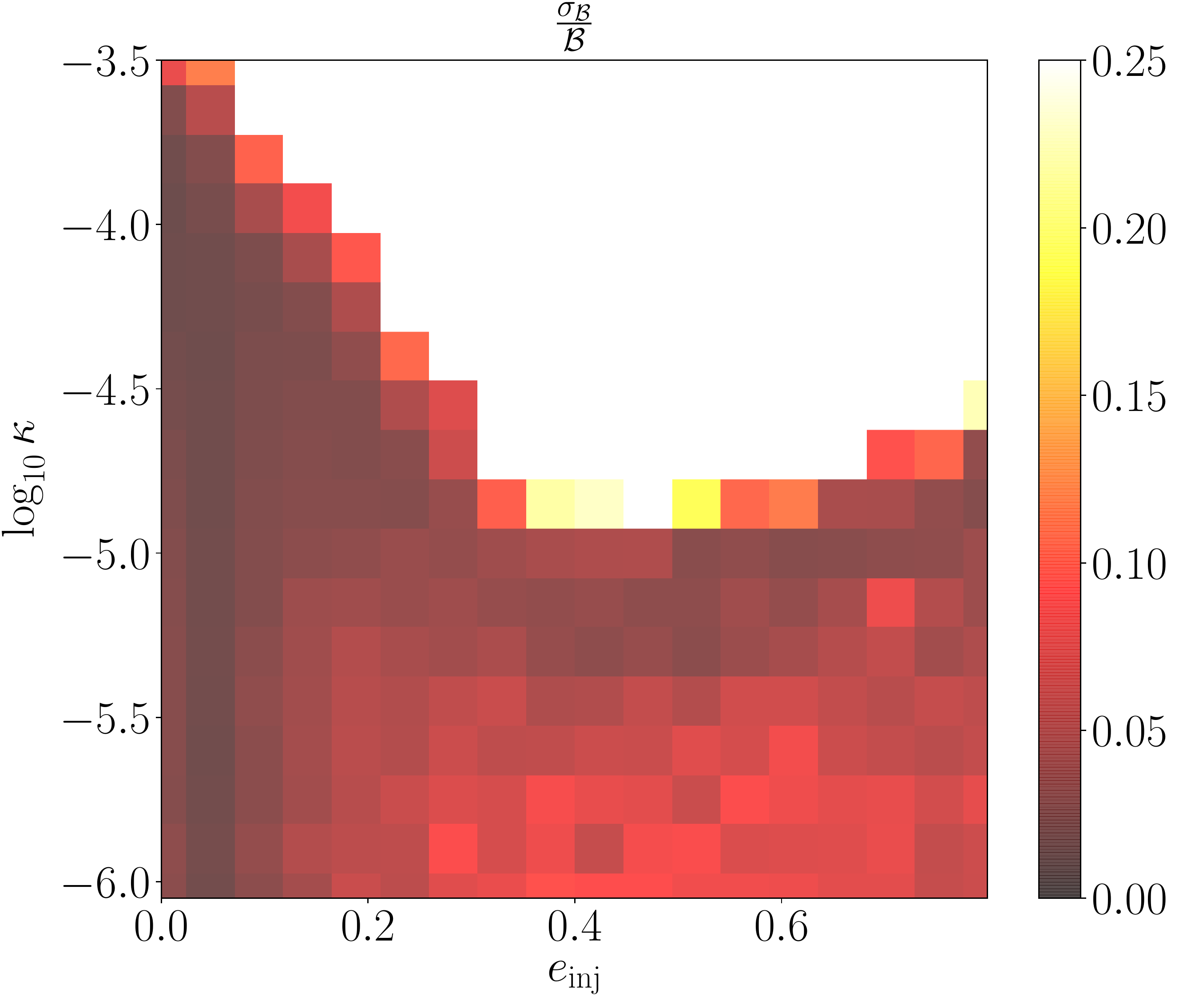}
\caption{\label{fig:Bayes_uncert} Standard deviation of the Bayes factor scaled by the Bayes factor as a function of the injected eccentricity and the non-GR parameter. For the regions of parameter space of interest, the uncertainty in Bayes factor is small enough and our reported Bayes factors are robust. }
\end{figure}
Here we seek to quantify the uncertainty in the Bayes factors shown throughout this work. The brute force method to calculate the uncertainty in the Bayes factor generated by an RJMCMC is to run identical algorithms with different seeds for the random number generator, which is used to draw jumps from the proposal distributions. This requires an  excessive amount of computational resources. Instead, here we use a metric developed in \cite{2015CQGra..32m5012C} to assess the error in the Bayes factor calculation. In that paper, the authors model the joint likelihood of observing the possible transitions of the RJMCMC, $N_{ij}$, where $N_{ij}$ are the number of transitions from state $i$ to state $j$ of the RJMCMC. This allows them to compute the variance of the Bayes factor. In terms of the standard deviation, this reads:
\begin{equation}
\sigma_{\mathcal{B}_{1,2}} = \mathcal{B}_{1,2}\left[ \frac{(N_1 - N_{12})}{N_1 N_{12}} + \frac{(N_2 - N_{21})}{N_2 N_{21}} \right]^{1/2} \, ,
\end{equation}
where $N_i$ is the number of iterations the RJMCMC algorithm spends in model $i$.

In Figure \ref{fig:Bayes_uncert} we show the values of the standard deviation of the Bayes factor scaled by the Bayes factor as a function of the non-GR parameter and the eccentricity of the signal. We have removed any value above $0.25$, as those regions with larger uncertainties have already been omitted in previous figures. We see that the standard deviation is mostly much below $0.25$, and the result shows that we can be confident in our reported Bayes factors in the regions of injected parameter space of interest. 

\section{Conclusions \& Future Work}
\label{sec:conc}
In this work we have derived and implemented corrections from Brans Dicke and Einstein-dilaton-Gauss-Bonnet theories of gravity into an eccentric waveform model which is valid in the moderate eccentricity regime. This is done by specifying the phase functions from these theories in terms of the orbital eccentricity which can than be trivially incorporated into the already existing TaylorF2e model \cite{my3PN, mydata}. We then carry out a comprehensive study of the data analysis implications of these alternative theories of gravity as a function of the eccentricity of the source. We set projected upper limits on the coupling parameter of BD and EdGB given an eccentric signal and employ a Bayes factor analysis exploring at what values of the injected alt theory parameter the non-GR model is favored. We also explore how non-trivial priors due to non-linear mapping between parameters can affect the results of these analyses. 

As a main result we find that regardless of the model selection or parameter estimation technique employed we find that moderately eccentric signals provide a more stringent constraint on the alternative theory of gravity than a circular signal with all other parameters held equal. When we assume the signal is consistent with GR we can set constraints on the coupling parameter about 10 times as stringent as in the circular case when the eccentricity is $\sim 0.4$. However, due to strong covariances with eccentricity at low eccentricities the constraint actually initially worsens for low eccentricity signals. The Bayes factor analysis is insensitive to this covariance and shows about a factor of $10^{3/2}$ increase in constraint of the alternative theory as the eccentricity is increased to $\sim 0.4$.

We also set projected constraints on the coupling parameters assuming an ideally eccentric signal with high SNR. We find a projected constrain for the EdGB parameter of $\sqrt{\alpha} < 0.5$~km which is an order of magnitude better than current constraints. A more conservative constraint is that given an eccentric detection which is otherwise comparable to current circular detections, we may be able to place a constraint that is a factor of 2 more stringent than provided by the circular signal. In the case of BD we are able to place a projected constraint of $\omega > 2700$ which is an order of magnitude less than the current constraint on $\omega$. Examination of Figure 5 of \cite{2019arXiv190807089M} which explores the constraint of $\omega$ for low eccentricity signals through a fisher analysis for several ground based detectors suggests that Cosmic Explorer or Einstein Telescope could provide a more stringent constraint than the current best especially for a moderately eccentric signal.

In deriving the eccentric corrections to the GW waveform we found that there is a simple mapping between the coupling parameters of the BD and EdGB theories of gravity in the way they affect the waveform. That is, the BD coupling parameter is related to the EdGB parameter by a constant factor. This simple mapping between the effects of alternative theories of gravity on the waveform suggests that a simple extension of the widely used ppE formalism exists for eccentric binaries. Given this work demonstrates that eccentric signals provide enhanced constraints on the alternative theories like EdGB and BD, if an extension to the PPE formalism does exist for eccentric signals its derivation would maximize the science that could be done with eccentric signals. 

This works underscores the importance of eccentric modeling and analysis. We have seen that eccentric signal not only increases our ability to validate GR but also lead to better measurement of source parameters \cite{mydata}. They also help complete our picture of how black hole binaries are formed due to their assembly being dependent of dynamic processes particular to certain astrophysical settings. We still require even more accurate models which are suitable for future detectors with the ability to incorporate the effects of several alternative theories of gravity to maximize the science potential of current and future detectors. 
\section*{Acknowledgments} 

B. M. was supported by the Joan L. Dalton Memorial Fellowship in Astronomy from Montclair State University. B.~M.~and N.~Y.~ also acknowledge support from NSF PHY-1759615 and NASA ROSES grant 80NSSC18K1352. We thank Travis Robson and Neil Cornish for very useful conversations. Computational efforts were performed on the Hyalite High Performance Computing system, which is supported by University Information Technology at Montana State University.

\bibliography{master}
\end{document}